\documentclass[a4paper,fleqn,usenatbib]{mnras}

\usepackage[T1]{fontenc}
\usepackage{ae,aecompl}

\usepackage{multirow} 
\usepackage{rotating}                   
\usepackage[acronym,nomain]{glossaries} 
\usepackage{lscape}

\usepackage{graphicx}	
\usepackage{amsmath}	
\usepackage{amssymb}	

\newacronym{los}{LOS}{line of sight}
\newacronym{blr}{BLR}{broad line region}
\newacronym{agn}{AGN}{active galactic nuclei}
\newacronym{hetgs}{HETGS}{High--Energy Transmission Grating Spectrometer}
\newacronym{dof}{dof}{degrees of freedom}
\newacronym{sed}{SED}{spectral energy distribution}
\newacronym{epic}{EPIC}{European Photon Imaging Camera}
\newacronym{ew}{EW}{equivalent width}
\newacronym{ufo}{UFO}{ultra--fast outflow}
\newacronym{rgs}{RGS}{Reflection Grating Spectrometer}
\newacronym{nls1}{NLS1}{Narrow Line Seyfert\,1}
\newacronym{gr}{GR}{general relativity}
\newacronym{smbh}{SMBH}{supermassive black hole}
\newacronym{isco}{ISCO}{innermost stable circular orbit}
\newacronym{cf}{CF}{covering fraction}
\newacronym{fwhm}{FWHM}{full width at half maximum}

\def\xmm{{\it XMM--Newton}}
\def\chandra{{\it Chandra}}

\def\mnras{MNRAS}       
\def\apj{ApJ}           
\def\apjs{ApJS}         
\def\aap{A\&A}          
\def\apjl{ApJ Letters}  
\def\araa{ARA\&A}       
\newcommand{\xr}{X--ray}
\newcommand{\kms}{km\,s$^{-1}$}
\newcommand{\ie}{i.e.}
\newcommand{\eg}{e.g.}

\newcommand{\kev}{keV}

\newcommand{\ev}{eV}

\newcommand{\ergps}{erg\,s$^{-1}$}

\newcommand{\fe}{Fe\,K$\alpha$}
\newcommand{\etal}{et al.}
\newcommand{\leqsim}{\raisebox{-0.6ex}{$\,\stackrel
        {\raisebox{-.2ex}{$\textstyle <$}}{\sim}\,$}}
\newcommand{\geqsim}{\raisebox{-0.6ex}{$\,\stackrel
        {\raisebox{-.2ex}{$\textstyle >$}}{\sim}\,$}}
\newcommand{\swi}{SWIFT\,J2127.4+5654}
\newcommand{\eso}{ESO\,323-G77}

\title[An ionised outflow in \eso] {
The ionised \xr\ outflowing torus in \eso : low--ionisation clumps confined by homogeneous warm absorbers
}

\author[M. Sanfrutos \etal]{
M.~Sanfrutos,$^{1}$\thanks{sanfrutoscm@cab.inta--csic.es} 
G.~Miniutti,$^{1}$
Y.~Krongold,$^{2}$ 
B.~Ag\'is-Gonz\'alez$^{1}$ 
and A.~L.~Longinotti$^{2, 3, 4}$
\\
$^{1}$Centro de Astrobiolog\'ia (CSIC--INTA), Dep. de Astrof\'isica; 
ESAC, PO Box 78, Villanueva de la Ca\~nada, E-28691 Madrid, Spain\\
$^{2}$Instituto de Astronom\'ia, Universidad Nacional Aut\'onoma de M\'exico, Apdo. 70-264, Cd. Universitaria, M\'exico DF 04510, M\'exico\\
$^{3}$European Space Astronomy Centre of ESA P.O. Box 78, Villanueva de la Ca\~nada, E-28691 Madrid, Spain\\
$^{4}$Catedr\'atica CONACYT - Instituto Nacional de Astrof\'isica, \'Optica y Electr\'onica, Luis E. Erro 1, Tonantzintla, Puebla, M\'exico, C.P. 72840
}

\date{Accepted 2015 December 17.  Received 2015 December 14; in original form 2015 May 8}

\pubyear{2015}

\begin{document}
\label{firstpage}
\pagerange{\pageref{firstpage}--\pageref{lastpage}}
\maketitle

\begin{abstract}
We report on the long-- and short--term \xr\ spectral analysis of the
polar--scattered Seyfert\,1.2 galaxy \eso, observed in three epochs
between 2006 and 2013 with \chandra\ and \xmm . Four high--resolution
\chandra\ observations give us a unique opportunity to study the
properties of the absorbers in detail, as well as their short
time--scale (days) variability. From the rich set of absorption
features seen in the \chandra\ data, we identify two warm absorbers
with column densities and ionisations that are consistent with being
constant on both short and long time--scales, suggesting that those
are the signature of a rather homogeneous and extended outflow. A
third absorber, ionised to a lesser degree, is also present and it
replaces the strictly neutral absorber that is ubiquitously inferred
from the \xr\ analysis of obscured Compton--thin sources. This
colder absorber appears to vary in column density on long
time--scales, suggesting a non--homogeneous absorber. Moreover, its
ionisation responds to the nuclear luminosity variations on
time--scales as short as a few days, indicating that the absorber is
in photoionisation equilibrium with the nuclear source on these
time--scales. All components are consistent with being co--spatial and
located between the inner and outer edges of the so--called dusty,
clumpy torus. Assuming co--spatiality, the three phases also share the
same pressure, suggesting that the warm\,/\,hot phases confine the
colder, most likely clumpy, medium. We discuss further the properties
of the outflow in comparison with the lower resolution \xmm\ data.
\end{abstract}



\section{Introduction}

\Gls{agn} ordinarily show \xr\ spectral variability on months to years
time--scales, which is often related to absorption phenomena
\citep[\eg][]{risaliti_var2002, miniutti2014, agis2014}. In many
cases, such long--term absorption variability can be associated with
the transit of dusty clouds in our \gls{los}, which reveals the
presence of a clumpy, dusty torus at relatively large spatial scales,
see \eg\ \cite{markowitz2014, agis2014}. 

In the last few years, various examples of absorption variability
within time--scales as short as hours or days have been reported, such
as in NGC\,4388, NGC\,4151, NGC\,1365, NGC\,7582, and \swi , as
reported in \cite{elvis2004}, \cite{puccetti2007}, \cite{Risaliti09},
\cite{bianchi2009}, and \cite{sanfrutos2013} respectively. As an
example, the in--depth study of the short time--scale absorption
variability in \swi\ reveals unambiguously the transit of a single
cloud in the \gls{los} to a fairly compact \xr\ source (few
gravitational radii in size). Usually, the short time--scale
absorption variability data are in good agreement with the existence
of a set of dense, cold clouds with characteristic column densities of
$10^{23}$ to $10^{24}$\,cm$^{-2}$, physical densities of $10^9$ to
$10^{11}$\,cm$^{-3}$ and velocities of the order of
$10^{3}$\,\kms\ orbiting the \xr\ source at radii of $10^3$ to
$10^4$\,$r_g$, where $r_g = GM/c^2$ is the gravitational radius for a
black hole of mass $M$. These properties suggest to identify the
obscuring clouds with the same clouds that are responsible for the
emission of broad optical\,/\,UV emission lines, \ie\ with clouds in
the \gls{blr}.

In the following we report results from four high--resolution
\chandra\ observations of \eso\ taken between April the 14$^{\rm th}$
and the 24$^{\rm th}$ 2010. \eso\ is a bright ($13.56$\,mag)
polar--scattered Seyfert\,1.2 galaxy \citep{veron2006} at $z=0.015$
\citep{dickens1986}. It was first classified as an \gls{agn} by
\cite{fairall1986}. The symmetry axis inclination of \eso\ is most 
likely of $\sim$\,$45^\circ$ with respect to our \gls{los} 
\citep{schmid2003}, intermediate between 
the characteristic inclination of Seyfert\,1 and Seyfert\,2 galaxies. Our
viewing angle is therefore likely grazing the edge of the obscuring matter,
namely the torus of the Unified model \citep{antonucci1993}. In order
to perform a more complete analysis and to compare the absorbers' properties at
different epochs, data from two high--quality \xmm\ observations are
also included from February the 7$^{\rm th}$ 2006 \citep{bailon2008}
and January the 17$^{\rm th}$ 2013 \citep{miniutti2014}.

\section{X--ray Observations}

\xmm\ first observed \eso\ on February the 7$^{\rm th}$ 2006 for a
total net exposure time of $\sim$\,$23$\,ks. Then, \chandra\ observed the
source on four occasions in April 2010 with the \gls{hetgs}: on the
14$^{\rm th}$ (ID: 11848, for a total net exposure time of
$\sim$\,$46$\,ks), on the 19$^{\rm th}$ (ID: 12139, $\sim$\,$60$\,ks), on
the 21$^{\rm st}$ (ID: 11849, $\sim$\,$118$\,ks) and on the 24$^{\rm th}$
(ID: 12204, $\sim$\,$67$\,ks). \xmm\ observed the source again on January
the 17$^{\rm th}$ 2013 for a total net exposure time of $\sim$\,$89$\,ks. Both \xmm\ observations (IDs: 0300240501 and 0694170101
respectively) were performed in ``{\small{Full Window}}'' mode with
the optical ``{\small{Thin}}'' filter applied. Standard data reduction was made
with the {\small{SAS v12.0.1}} software for \xmm\ and with
the {\small{CIAO v4.5}} software for \chandra. Observation--dependent
redistribution matrices and ancillary responses were
generated as standard for every data set. Spectral analysis was
performed using the {\small{XSPEC v12.8.1}} software
\citep{arnaud_xspec1996}.

\xmm\ \gls{epic} source products were extracted from source--centred
circular regions, and the corresponding background ones were estimated
from source--free nearby areas. For the sake of simplicity, and after
having checked the good agreement among the pn, MOS1 and MOS2 data,
only \gls{epic}-pn spectra in the 0.5--10\,\kev\ band are used in this
study. As for \chandra, we use the MEG data in the 1.2--7\,\kev\ band,
and the HEG ones in the 1.4--9\,\kev . Outside these energy bands, the
high--resolution spectra are background--dominated. We used these
spectra in two forms: (i) separately when interested on the short
time--scale variability of the absorbers, and (ii) merged into one
single $\sim$\,$291$\,ks spectrum for each detector, representative of the
broad--band \xr\ continuum time--averaged over 10 days. The
\xmm\ spectra have been regrouped so that each bin contains 25 counts,
while the \chandra\ spectra have been grouped to 4 channels per bin,
and we use the $\chi^2$ and $C$--statistic \citep{cash1979} for the \xmm\ and
\chandra\ spectral analysis respectively. Uncertainties correspond to
the $90\%$ confidence level for one interesting parameter, except if
otherwise specified. Whenever fluxes were needed to be converted into
luminosities, we have assumed a $\Lambda$CDM cosmology with
H$_0=70$\,\kms\,Mpc$^{-1}$, $\Omega_\Lambda= 0.73$, and $\Omega_M =
0.27$.

\begin{figure}
\begin{center}
\includegraphics[height=0.42\textwidth,angle=-90]{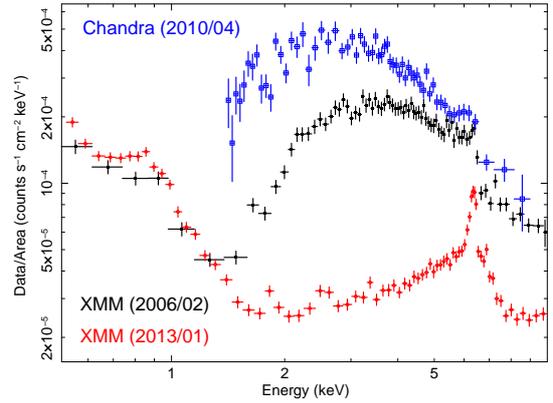}
\caption{\label{fig:xmmchanxmm}Long--term \xr\ spectral variability of
  \eso\ as observed with \xmm\ and \chandra . Only pn and HEG data
  respectively are shown for clarity. Data have been normalised to
  each detector effective area, and they have been re--binned for
  visual clarity. }
\end{center}
\end{figure}

The \xr\ spectra of the  \xmm\ and \chandra\ observations
can be seen in Fig.\,\ref{fig:xmmchanxmm}. In order to facilitate
comparison among them, all spectra have been divided by the effective
area of each detector. Spectral variability on long time--scales is
clearly present with absorption first decreasing between the first
\xmm\ observation (2006/02) and the \chandra\ one (2010/04) and then
increasing significantly between the \chandra\ observation (2010/04)
and the second \xmm\ one (2013/01), as already studied in detail
in our previous work on \eso\ \citep{miniutti2014}.

\section{The time--averaged 2010 \chandra\ spectrum}

We start our analysis by considering the merged MEG and HEG data from
\chandra, \ie\ we consider the time--averaged, high--resolution
\chandra\ data that are representative of the spectrum between April
the 14$^{\rm th}$ and the 24$^{\rm th}$ 2010. 

Based on our previous analysis of the source \citep{miniutti2014}, we
consider a baseline model comprising Galactic absorption
\citep{kalberla2005}, a power law \xr\ continuum, a reflection
continuum \citep{nandra_pexmon2007} from neutral matter (with Solar
abundances and inclination fixed to an intermediate value of
$45^\circ$), and a scattered soft \xr\ power law typical of obscured
\gls{agn} \citep{matt2013}. The photon indices of the nuclear
continuum and of the soft scattered component are forced to be the
same, while their normalisations are free to vary independently. The
reflection model intensity is set by the so--called reflection
fraction $R$, with the geometrical meaning that $R=1$ corresponds to
the reflector covering half of the sky as seen by the irradiating
source. The reflection model is convolved with a Gaussian kernel to
account for any width of the associated emission lines (mainly Fe
K$\alpha$). As for the absorbing systems we include, as a first
approximation, a neutral absorber fully covering the nuclear \xr\ continuum. A
constant is introduced to account for calibration uncertainties
between the two detectors.

The best--fitting baseline model produces a statistical result of $C =
2190 $ for 1168 \gls{dof}. The photon index is $\Gamma
= 1.85 \pm 0.06$, and the \xr\ continuum is absorbed by a column
density of N$_{\rm H} = (3.3 \pm 0.2) \times 10^{22}$\,cm$^{-2}$. The
soft scattered power law has a normalisation that is about 4 per cent
that of the nuclear \xr\ continuum. The reflection fraction of the
reflection model is $R = 0.27 \pm 0.12$ and replacing the reflection
model with a simple Gaussian emission line at $\sim$\,$6.4$\,\kev\  gives a
line energy of $6.39 \pm 0.02$\,\kev\  with \gls{ew} of $50\pm
10$\,eV. The width of the line is in the range of $3-30$\,\ev\ and
corresponds to a \gls{fwhm} $\sim$\,$330-3300$\,\kms , consistent with an
origin in the \gls{blr} or further out (\eg\ the
so--called torus). In our analysis we fix the width of the
Gaussian kernel applied to the neutral reflection model to an
intermediate value of $10$\,\ev.

The best--fitting continuum model reveals the presence of a series of
relatively strong absorption lines in the $1.2-2.7$\,\kev\  range and
around $7$\,\kev . We then add a series of Gaussian absorption lines to our
best--fitting model. Each line is in principle associated with three
free parameters, namely rest--frame energy, width, and
intensity. After a few initial tests, we find that the width of most
absorption lines can not be well constrained by the data. In order to gain
some insigth on the typical line width, we select two of the strongest
soft \xr\ lines (at $\sim$\,$1.87$\,\kev\ and $\sim$\,$2.02$\,\kev\ at the
galaxy redshift) and the two Fe absorption lines (at $\sim$\,$6.73$\,\kev\ 
and $\sim$\,$6.99$\,\kev ), and we fit those lines with free Gaussian
width. The two soft \xr\ lines are both consistent with $\sigma =
10\pm 3$\,\ev, while the two highly ionised Fe lines have 
$\sigma$\,$\leq$\,$7$\,\ev . Hence, in the subsequent analysis, we fix 
the width of all Gaussian absorption lines to $10$\,\ev, except that of 
the two highly ionised Fe lines which is instead fixed at $1$\,\ev. Each 
Gaussian then only contributes with two free parameters (rest--frame 
energy and intensity). The relatively large width of the soft \xr\ lines 
is likely an indication of turbulence and\,/\,or of a contribution of
different gas phases with different velocities to most lines.

\begin{table*} 
\caption{\label{tab:abs_lines}Absorption lines detected with Gaussian
  models in the time--averaged \chandra\ spectra from the MEG and HEG
  detectors.  The phase label (first column) refers to the phase of
  the gas that is likely responsible for the specific absorption line
  and it is coded as follows: h~=~high--ionisation warm absorber (with
  typical temperature in the range of $10^6-10^7$\,K),
  l~=~low--ionisation warm absorber ($10^5-10^6$\,K), c~=~cold
  absorber ($10^4-10^5$\,K). The last column is the statistical
  improvement associated with the corresponding Gaussian model. We
  gather all the atomic transitions data from the {\small{AtomDB}}
  \citep{smith_apec2001}.}
\begin{center}
\begin{tabular}{l l c c c r r r}
\hline\hline 
Phase & ID & Transition & E$_{\rm lab}$\,(\kev)\,/\,$\lambda_{\rm lab}$\,(\r{A}) & E$_{\rm restframe}$\,(\kev)& $-$EW\,(eV)& v$_{\rm outflow}$\,(\kms) & $\Delta\,C$ \\
\hline
l$+$c & Ne\,\textsc{x}   & $1s   \rightarrow 4p$           & $1.2770$\,/\,$9.708$ & $1.283\pm 0.002$ & $20\pm 4$ & $1400\pm 450$ & $40$  \\
l$+$c & Ne\,\textsc{x}   & $1s   \rightarrow 5p$           & $1.3077$\,/\,$9.481$ & $1.314\pm 0.002$ & $15\pm 4$ & $1450\pm 450$ & $18$  \\
l$+$c & Mg\,\textsc{xi}  & $1s^2 \rightarrow 1s 2p$        & $1.3522$\,/\,$9.169$ & $1.359\pm 0.002$ & $9 \pm 5$ & $1500\pm 450$ & $11$  \\
l$+$c & Mg\,\textsc{xii} & $1s   \rightarrow 2p$           & $1.4723$\,/\,$8.421$ & $1.480\pm 0.002$ & $15\pm 3$ & $1550\pm 400$ & $71$  \\
c & Si\,\textsc{viii}    & $2p^3 \rightarrow 1s 2s^2 2p^4$ & $1.7715$\,/\,$6.999$ & $1.777\pm 0.002$ & $13\pm 3$ & $950 \pm 300$ & $28$  \\
c & Si\,\textsc{ix}      & $2p^2 \rightarrow 1s 2s^2 2p^3$ & $1.7909$\,/\,$6.923$ & $1.797\pm 0.002$ & $12\pm 2$ & $1000\pm 350$ & $27$  \\
c & Si\,\textsc{x}       & $2p   \rightarrow 1s 2s^2 2p^2$ & $1.8084$\,/\,$6.856$ & $1.816\pm 0.002$ & $11\pm 3$ & $1250\pm 350$ & $24$  \\
l & Si\,\textsc{xiii}    & $1s^2 \rightarrow 1s 2p$        & $1.8650$\,/\,$6.648$ & $1.875\pm 0.002$ & $12\pm 2$ & $1600\pm 300$ & $63$  \\
l$+$h & Si\,\textsc{xiv} & $1s   \rightarrow 2p$           & $2.0056$\,/\,$6.182$ & $2.017\pm 0.002$ & $18\pm 2$ & $1700\pm 300$ & $181$ \\
l & Si\,\textsc{xiv}     & $1s   \rightarrow 3p$           & $2.3765$\,/\,$5.217$ & $2.388\pm 0.003$ & $14\pm 3$ & $1450\pm 350$ & $25$  \\
l & S\,\textsc{xv}       & $1s^2 \rightarrow 1s 2p$        & $2.4605$\,/\,$5.039$ & $2.473\pm 0.003$ & $10\pm 3$ & $1500\pm 400$ & $17$  \\
l$+$h & S\,\textsc{xvi}  & $1s   \rightarrow 2p$           & $2.6218$\,/\,$4.729$ & $2.632\pm 0.003$ & $11\pm 3$ & $1150\pm 350$ & $28$  \\
h & Fe\,\textsc{xxv}     & $1s^2 \rightarrow 1s 2p$        & $6.7019$\,/\,$1.850$ & $6.73 \pm 0.01$  & $49\pm 7$ & $1250\pm 450$ & $41$  \\
h & Fe\,\textsc{xxvi}    & $1s   \rightarrow 2p$           & $6.9650$\,/\,$1.780$ & $6.99 \pm 0.01$  & $51\pm 9$ & $1050\pm 450$ & $36$  \\
\hline\hline
\end{tabular}
\end{center}
\end{table*}

We detect a total of 14 absorption lines, each producing an
improvement of $\Delta C$\,$\geq$\,$9.2$, \ie\ each line is associated with a
statistical significance larger than $\sim$\,$99$~per cent for the two
free parameters. The final statistical result is of $C = 1580$ for 1140
\gls{dof}. In Table\,\ref{tab:abs_lines}, we report the best--fitting
parameters of the Gaussian absorption lines, as well as the
corresponding identification and inferred outflow velocity. Errors on
the lines parameters are computed using the {\small{STEPPAR}} command in
{\small{XSPEC}}. The $\Delta C$ improvement for each individual line is
computed by removing the line under inspection from the best--fitting
model and by re--fitting the data reaching a new best--fit to be
compared with the former.

As shown in Table\,\ref{tab:abs_lines}, we can identify three main
groups of absorption lines, likely associated with three different
ionisation states. The highest ionisation phase (h) is associated
with the Fe\,{\textsc{xxv}} and Fe\,{\textsc{xxvi}} absorption
lines, with possible contributions to Si\,{\textsc{xiv}} and
S\,{\textsc{xvi}}. An intermediate ionisation state (l) is probed mainly
by Si\,{\textsc{xiii}}, Si\,{\textsc{xiv}} and S\,{\textsc{xv}}
with other possible contribution at Ne\,{\textsc{x}},
Mg\,{\textsc{xi}} and Mg\,{\textsc{xii}}, as well as
at S\,{\textsc{xvi}}. Finally, a series of Si absorption lines
(Si\,{\textsc{viii-x}}) is associated with a
colder phase (c), which may also contribute, together with the
intermediate ionisation phase, at Ne\,{\textsc{x}},
Mg\,{\textsc{xi}} and Mg\,{\textsc{xii}}.

According to our lines ID in Table\,\ref{tab:abs_lines}, all phases have similar
outflow velocities (of the order of $1000-2000$\,\kms), although
excluding lines that may be associated with more than one single
phase, the high ionisation (h) and the cold phase (c) may be slighly slower
than the intermediate zone (l).

\subsection{A global absorption model}
\label{sec:global}

\begin{figure}
\begin{center}
\includegraphics[width=0.32\textwidth,height=0.42\textwidth,angle=-90]{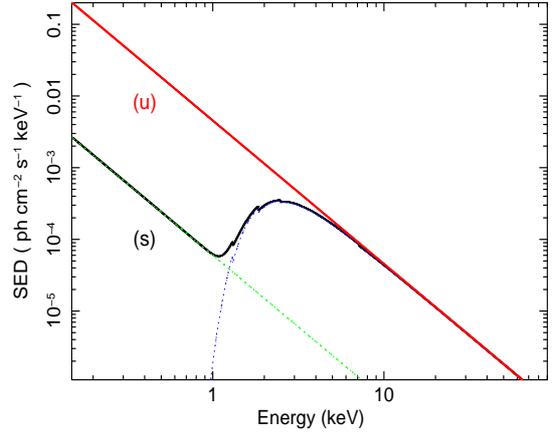}
\caption{The unabsorbed (u) and soft--X--rays--absorbed (s)
  \glspl{sed} used in this work are shown as solid lines (upper and
  lower lines respectively) in a restricted X--ray regime. The overall
  \glspl{sed} can be obtained with a simple extrapolation between
  1\,Ry and infinity. In the (s) case the two main spectral components
  are also shown, namely the soft scattered power law and the absorbed
  nuclear continuum which dominate the spectrum below and above 
  $\sim$\,$1$\,\kev\ respectively.}
\label{fig:sed}
\end{center}
\end{figure}

In order to reproduce the absorption features that are present in the
data, we start our analysis by applying two ionised absorbers to the
nuclear continuum, which is already absorbed by a neutral column of
$\sim$\,$3\times 10^{22}$\,cm$^{-2}$. We use the photoionisation code
{\small{PHASE}}, developed by
\cite{krongold_phase2003}. {\small{PHASE}} assumes a simple geometry
that consists of a central source emitting an ionising continuum with
clouds of gas intercepting the \gls{los}, in a plane parallel
approximation. The ionisation balance is calculated using
{\small{CLOUDY}} \citep[last described in][]{ferland13_cloudy}. In
this case, we use a simple power law \gls{sed} with photon index
$\Gamma = 2$ over the whole Lyman continuum. The \gls{sed} is
normalised to the historical 2--10\,\kev\  luminosity of
\eso\ \citep{miniutti2014}. 
This \gls{sed} is shown in red in Fig.\,\ref{fig:sed}, 
and denoted by (u) for \emph{unabsorbed}. 
We also include in the same figure another \gls{sed} 
that will be defined and used later in this Section. 
The parameters of the code are (1) the
ionisation parameter defined as $U = Q~(4 \pi n R^2 c)^{-1}$
\citep{netzer2008}, where $Q$ is the photon rate integrated over the
entire Lyman continuum, $n$ is the gas number density and $R$ the gas
distance from the nuclear source of photons, (2) the equivalent
hydrogen column density, (3) the outflow velocity, and (4) the
internal microturbulent velocity. The electron temperature in the
models presented here corresponds to the photoionisation equilibrium
of the gas.

\begin{figure*}
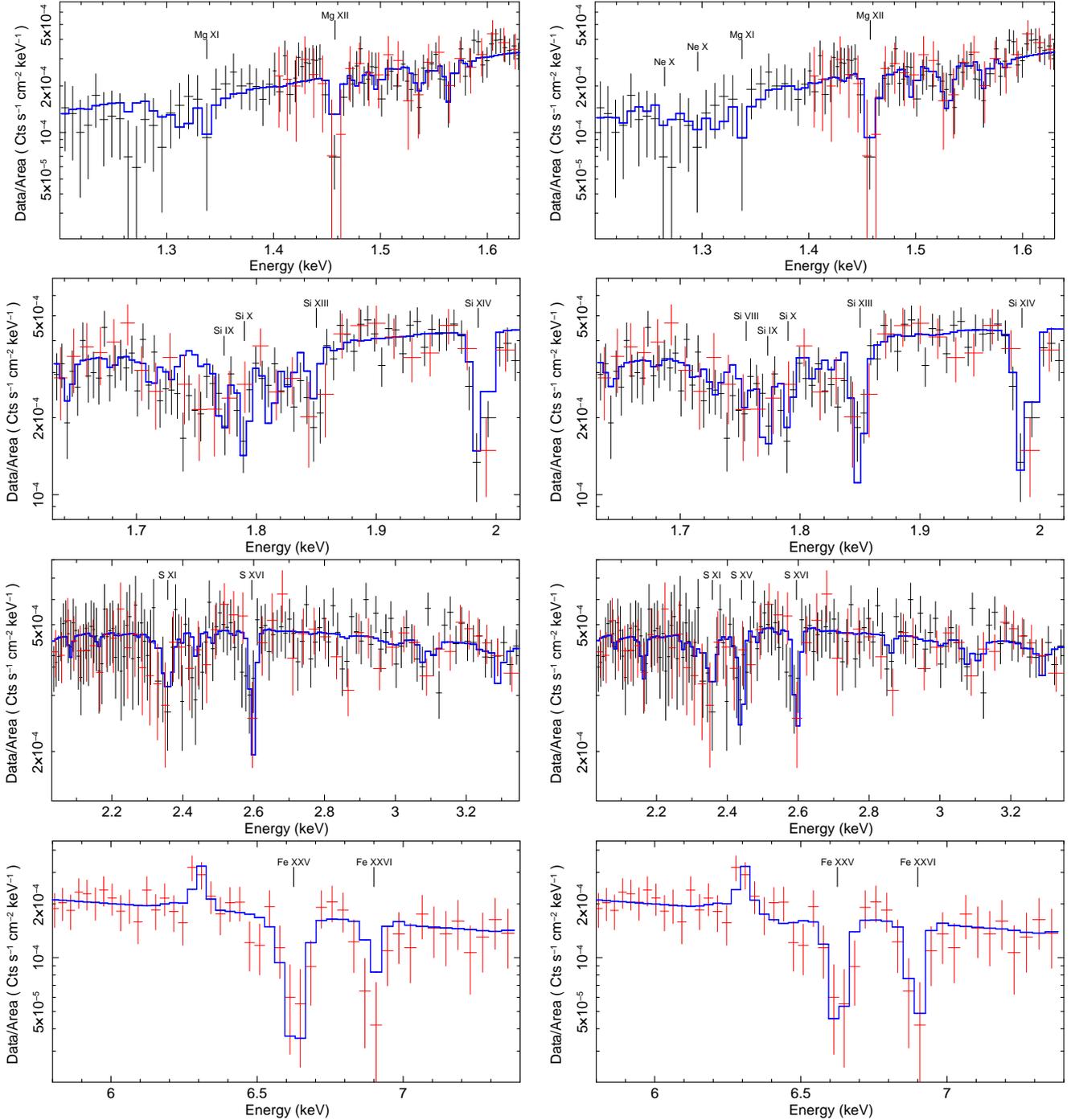

\begin{center}

\mbox{\includegraphics[width=0.26\textwidth,angle=-90]{Mg2phases.eps}}
\mbox{\includegraphics[width=0.26\textwidth,angle=-90]{Mg3phases.eps}}

{\vspace{0.0cm}}

\mbox{\includegraphics[width=0.26\textwidth,angle=-90]{Si2phases.eps}}
\mbox{\includegraphics[width=0.26\textwidth,angle=-90]{Si3phases.eps}}

{\vspace{0.0cm}}

\mbox{\includegraphics[width=0.26\textwidth,angle=-90]{S2phases.eps}}
\mbox{\includegraphics[width=0.26\textwidth,angle=-90]{S3phases.eps}}

{\vspace{0.0cm}}

\mbox{\includegraphics[width=0.26\textwidth,angle=-90]{Fe2phases.eps}}
\mbox{\includegraphics[width=0.26\textwidth,angle=-90]{Fe3phases.eps}}

\caption{We show the area--corrected MEG (black) and HEG (red) data
  and best--fitting models in four energy bands (from top to bottom
  they are: 1.2--1.63\,\kev , 1.63--2.05\,\kev , 2.05--3.35\,\kev ,
  and 5.8--7.4\,\kev ). We omit the intermediate 3.35--5.8\,\kev\ band
  as no strong lines are seen in this range.  The left panels refer to
  Model\,1 in Table\,\ref{tab:chandrafits} (namely two ionised and one
  neutral absorbers), and the right panels refer to Model\,2 (three
  ionised absorbers). 
  The best--fitting model (solid line) is only that convolved with the MEG response in the three top panels, 
  while only HEG data and model are shown in the bottom panels to ensure simplicity and visual clarity.}
\label{fig:chandrafits}
\end{center}
\end{figure*}

The two ionised absorbers provide a large statistical improvement with
respect to the baseline continuum model reaching $C = 1591$ for 1160
\gls{dof} (to be compared with $C= 2190$ for 1168) and the
best--fitting parameters are reported in Table\,\ref{tab:chandrafits}
(Model\,1).  The data and best--fitting model are shown in the left
panels of Fig.\,\ref{fig:chandrafits}. Although the model represents a
significant statistical improvement, some of the absorption lines we
detect with Gaussian models (see Table\,\ref{tab:abs_lines}) are only
poorly reproduced. This is particularily true for some Si lines
between 1.7\,\kev\ and 1.8\,\kev\ (observed--frame) as seen in the
second--left panel of Fig.\,\ref{fig:chandrafits}, for the
S\,\textsc{xv} line around 2.4\,\kev\ (third--left panel), and for the
Fe\,\textsc{xxvi} line that is only poorly accounted for (bottom--left
panel). Some residuals are also left at the softest \xr\ energies
(see top--left panel of Fig.\,\ref{fig:chandrafits}), although they
seem to have lower significance.

\begin{center}
\begin{table*}

\caption{\label{tab:chandrafits} Best--fitting parameters for the
  time--averaged MEG and HEG \chandra\ data. Column densities are
  expressed in cm$^{-2}$, the ionisation parameter $U$ is
  dimensionless by definition, while $\xi$ is in units of
  erg\,cm\,s$^{-1}$. The different columns are characterised by the
  different models for the absorbers. The symbol $n$ represents a
  strictly neutral absorber, $u$ and $s$ represent ionised absorbers
  modelled with the unabsorbed (u) or soft--absorbed (s) \gls{sed} in
  {\small{PHASE}}, and $i$ an ionised absorber modelled with the
  {\small{XSTAR}}--based model {\small{ZXIPCF}}. The overscript $^f$
  means that the parameter has been fixed, while the symbol $p$
  indicates that it reached the model upper\,/\,lower limit respectively.}
\begin{center}
\begin{tabular}{c c c c c c c }
\hline
   &   &   & Model\,1 & Model\,2 & Model\,3 & Model\,4 \\
   &   &   & u$\times$u$\times$n   & u$\times$u$\times$u & s$\times$s$\times$u & i$\times$i$\times$i\\
\hline\hline
& & $\Gamma$ & $1.85 \pm 0.06$ & $1.96\pm 0.05$ & $1.95\pm 0.05$ & $2.15\pm 0.07$\\
\hline   
\multirow{10}{*}{\begin{turn}{90}Warm abs.\end{turn}} & \multirow{5}{*}{\begin{turn}{90}h--phase\end{turn}}  & log\,${\rm U}$ & $1.5\pm 0.1$ & $1.9\pm 0.2$ & $0.0\pm 0.2$ & $-$ \\
& & log\,${\rm \xi}$ & $-$ & $-$ & $-$ & $4.1\pm 0.1$ \\
& & log\,N$_{\rm H}$ & $23.2 \pm 0.1$ & $23.4\pm 0.2$ & $23.3\pm 0.2$ & $23.7\pm 0.2$\\
& & v$_{\rm turb}$ & $600-900^p$ & $600-900^p$ & $600-900^p$ & $200^{f}$\\
& & v$_{\rm outflow}$ & $1500\pm 200$ & $1100\pm 200$ & $1200\pm 200$ & $800\pm 200$\\
\cline{2-7}
& \multirow{5}{*}{\begin{turn}{90}l--phase\end{turn}}  & log\,${\rm U}$ & $-0.3\pm 0.1$ & $0.5\pm 0.1$ & $-1.0\pm 0.1$ & $-$\\
& & log\,${\rm \xi}$ & $-$ & $-$ & $-$ & $2.7\pm 0.2$ \\
& & log\,N$_{\rm H}$ & $22.50 \pm 0.08$ & $22.41\pm 0.09$ & $22.49\pm 0.08$ & $22.5\pm 0.1$\\
& & v$_{\rm turb}$ & $500-900^p$ & $600-900^p$ & $600-900^p$ & $200^{f}$ \\
& & v$_{\rm outflow}$ & $1400\pm 200$ & $1750\pm 200$ & $1700\pm 200$ & $1850\pm 200$\\
\hline 
\multirow{5}{*}{\begin{turn}{90}Cold abs.\end{turn}} & \multirow{5}{*}{\begin{turn}{90}c--phase\end{turn}} & log\,${\rm U}$ & -- & $-0.45\pm 0.05$ & $-0.56\pm 0.05$ & $-$\\
& & log\,${\rm \xi}$ & $-$ & $-$ & $-$ & $1.6\pm 0.2$ \\
& & log\,N$_{\rm H}$ & $22.20\pm 0.06$ & $22.70\pm 0.04$ & $22.59\pm 0.05$ & $22.9\pm 0.2$\\
& & v$_{\rm turb}$ & -- & $600\pm 200$ & $600\pm 200$ & $200^{f}$ \\
& & v$_{\rm outflow}$ & -- & $1400\pm 300$ & $1300\pm 300$ & $1500\pm 300$\\
\hline\hline
& & C\,/\,\gls{dof} & 1591\,/\,1160 & 1433\,/\,1157 & 1413\,/\,1157 & 1700\,/\,1160 \\
\hline\hline
\end{tabular}
\end{center}
\end{table*}
\end{center}

In our model, the series of Si lines around 1.8\,\kev\ are all due to
the low--ionisation l--phase ($\log U$\,$\sim$\,$-0.3$, see Model\,1 in
Table\,\ref{tab:chandrafits}), while the strong Si\,\textsc{xiv} and
S\,\textsc{xvi} lines are exclusively produced by the high--ionisation
h--phase together with the Fe lines ($\log U$\,$\sim$\,$1.5$). Increasing
the ionisation of this hotter phase to better reproduce the
Fe\,\textsc{xxvi} line decreases the strength of the Si\,\textsc{xiv}
and S\,\textsc{xvi} ones, and worsen the fitting statistics because
the l--phase is not hot enough to contribute there. This would be
possible only increasing as well the ionisation of the l--phase, but
then the other Si lines would not be reproduced, as they are
associated with lower ionisation.

The only sensible solution seems that of introducing a cold phase
which may account for the Si lines around 1.8\,\kev\  allowing the
ionisation of the other two components to increase and to better
reproduce the Si\,\textsc{xiv}, S\,\textsc{xv}, S\,\textsc{xvi}, and
Fe\,\textsc{xxvi} lines. It would obviously be possible to simply
introduce a third {\small{PHASE}} component. However, the ionisation
is likely to be low, so that we consider the possibility of
replacing the cold asorber that characterises our continuum model with
a low--ionisation absorber. Hence we replace the neutral absorber (so
far modelled with the {\small{ZPHABS}} model in {\small{XSPEC}}) with
a third {\small{PHASE}} component.

\begin{figure*}
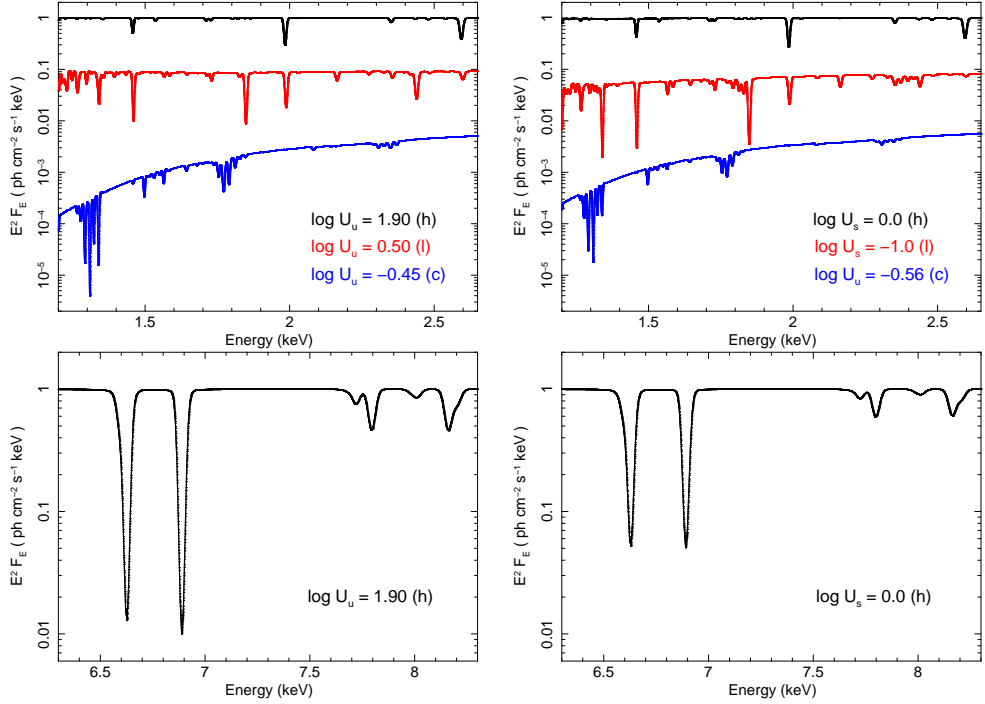

\begin{center}

\mbox{\includegraphics[width=0.26\textwidth,angle=-90]{3phases_softnew.eps}}
\mbox{\includegraphics[width=0.26\textwidth,angle=-90]{3phases_soft_ssunew.eps}}

{\vspace{0.0cm}}

\mbox{\includegraphics[width=0.26\textwidth,angle=-90]{3phases_hardnew.eps}}
\mbox{\includegraphics[width=0.26\textwidth,angle=-90]{3phases_hard_ssunew.eps}}

\caption{In the top panels we show the best--fitting models for the
  three {\small{PHASE}} components in the soft \xr\ band between
  $1.2$\,\kev\ and $2.65$\,\kev\ (where the strongest features are seen). 
  The bottom panels show the Fe\,K region, where the only significant
  contribution is from the highly--ionised component. 
  The models are all applied to a power law with $\Gamma = 2$ and are
  rescaled by factors of 10 in order to avoid confusion. 
  The left panels refer to
  Model\,2 in Table\,\ref{tab:chandrafits} (u$\times$u$\times$u 
  configuration), where all absorbers see the same, unabsorbed \gls{sed} (u). 
  The right panels are for Model\,3 (s$\times$s$\times$u 
  configuration), where the h-- and l--phases see a \gls{sed} that is 
  absorbed by a neutral column of $3 \times 10^{22}$\,cm$^{-2}$ (s), 
  while the c--phase sees the unabsorbed \gls{sed} (u). 
  As can be seen,
  the two models differ slightly in some of the line ratios, while the
  general structure is the same. In particular, the l--phase is always
  responsible for higher--ionisation lines than the c--phase despite
  the nominally lower $U$ obtained with Model\,3 (right panels).}
\label{fig:3phases_models}
\end{center}
\end{figure*}

Replacing the neutral absorber with a third {\small{PHASE}} component
produces a statistically significant improvement, and the best--fit
reaches $C = 1433$ for 1157 \gls{dof} (to be compared with $C = 1591$ for
1160 \gls{dof}, obtained with two ionised and one strictly neutral
absorber). The best--fitting parameters are reported in
Table\,\ref{tab:chandrafits} (Model\,2). The data and best--fitting
model are shown in the right panels of Fig.\,\ref{fig:chandrafits}. The
best--fitting models are shown in the left panels of
Fig.\,\ref{fig:3phases_models} in the two energy bands where the most
relevant features are imprinted (a soft \xr\ band up to
$\sim$\,$2.7$\,\kev\  is shown in the upper panel, and the Fe\,K region in the
lower).

As expected, the ionisation of the two absorbers of our previous model
significantly increases. The h--phase is now characterised by 
$\log U$\,$\sim$\,$1.9$, while the l--phase has $\log U$\,$\sim$$\,0.5$. 
Now, the Si\,\textsc{xiv} and S\,\textsc{xvi} lines are well reproduced 
with contribution from both absorbers, while the high--ionisation phase
accounts very well for both the Fe\,\textsc{xxv} and Fe\,\textsc{xxvi}
lines. The l--phase also accounts better for the S\,\textsc{xv}
absorption feature around 2.4\,\kev .  On the other hand, the new,
colder absorber ($\log U$\,$\sim$\,$-0.45$) accounts for the Si complex
around $1.8$\,\kev\ (and also slightly improves the fit at the lowest
enegies, where Ne and Mg lines are detected, see
Table\,\ref{tab:abs_lines}). All the improvements of Model\,2 with
respect to Model\,1 can be seen in the comparison between left
(Model\,1) and right (Model\,2) panels of Fig.\,\ref{fig:chandrafits}.

We measure a non--zero velocity for all phases. The outflow velocities
are of $1100\pm 200$\,\kms\ for the h--phase, of $\sim$\,$1750\pm
200$\,\kms\ for the l--phase, and of $1400\pm 300$\,\kms\ for the
coldest absorber (c--phase). The l--phase appears to be faster than
the other two. This is consistent with the results reported in
Table\,\ref{tab:abs_lines} where it was shown that lines associated
with the h--phase only (the Fe lines) and those associated with the
c--phase only (the Si\,\textsc{viii} to Si\,\textsc{x} lines) have
velocities in the range of $900-1300$\,\kms , while lines associated
with the l--phase only (such as Si\,\textsc{xiii} and S\,\textsc{xv})
have marginally higher velocity ($1300-1800$\,\kms ). As for the
turbulent velocities, the only one that can be constrained is that of
the coldest component which is v$_{\rm{turb}}^{\rm{(c)}} = 600 \pm 200
$\,\kms . Only lower limits are obtained for the other two phases with
$600$\,\kms\,$\leq$\,${\rm{v}}_{\rm{turb}}^{\rm{(l,h)}}$\,$\leq$\,$900$\,\kms\ 
(note that the {\small{PHASE}} code only allows for v$_{\rm{turb}}$\,$\leq$\,$900$\,\kms ).

As mentioned, we have so far used a common \gls{sed} (a simple power
law with photon index $2$ over the whole Lyman continuum) for the
three absorbers. This \gls{sed} is represented with the symbol $u$ for
{\emph{unabsorbed}} in Table\,\ref{tab:chandrafits}. However, as can be
seen in the top--left panel of Fig.\,\ref{fig:3phases_models}, the cold
component affects the spectral shape significantly in the soft
\xr s. If the cold absorber is located closer to the nuclear source
than the other phases, the h-- and l--phase would see a different,
absorbed \gls{sed}. To explore the possible effects of such a
scenario, we introduce a new \gls{sed}, namely we consider the same
nuclear power law with $\Gamma = 2$ as before, but absorbed by a
column density of $3\times 10^{22}$\,cm$^{-2}$ of cold gas
(representative of the effect of the c--phase). In the soft \xr s
(and below), this new \gls{sed} is therefore dominated by the soft,
scattered power law which is always present. The luminosity of the
soft power law is $L_{\rm{0.5-2}} = 6.7\times 10^{40}$\,\ergps\ 
while, as already mentioned, that of the nuclear continuum is the
historical average for \eso\ ($L_{\rm{2-10}} = 5.8\times
10^{42}$\,\ergps ). Both luminosities are from
\cite{miniutti2014}. This absorbed \gls{sed} is 
represented with the symbol $s$ for {\emph{soft--X--rays--absorbed}} 
in Table\,\ref{tab:chandrafits}, 
and is shown in black in Fig.\,\ref{fig:sed} in order to ease comparison 
with the unabsorbed \gls{sed}. 
The relation between the
integrated photon rate $Q$ for the unabsorbed \gls{sed} (u) and the new,
soft--\xr --absorbed \gls{sed} (s) is $Q_{\rm u}$\,$\simeq$\,$50\,Q_{\rm s}$. 
The introduction of this new model enables us to
explore the possibility that the c--phase is closer to the nuclear
source than the other two (or at least co--spatial with them), thus
reducing the irradiating flux on the h-- and l--phases. As the two
warm\,/\,hot phases see the absorbed \gls{sed}, while the c--phase still
sees the unabsorbed one, the model is called s$\times$s$\times$u, as
opposed to the case in which all phases see the same unabsorbed
\gls{sed} which is called u$\times$u$\times$u.

Repeating the analysis in the s$\times$s$\times$u configuration
produces a relatively marginal improvement and gives $C=1413$ for the
same number of \gls{dof} as in the u$\times$u$\times$u explored above
(which gave $C=1433$) and results are reported in
Table\,\ref{tab:chandrafits} as Model\,3. We believe that the $\Delta
C$ between the two models is not sufficiently large to prefer one
solution over the other on firm statistical grounds, but we take it as
an indication that the c--phase is either more internal or at least
co--spatial with the other two, rather than more external.

Note that the ionisation parameter of the two higher--ionisation warm
absorbers drops significantly with $\Delta \log U = \log U_u - \log
U_s =1.9 \pm 0.4$ for the h--phase and $1.5 \pm 0.2$ for the
l--phase. This mostly reflects the change in ionising photon rate $Q$
due to the reduction of the \gls{sed} at soft \xr s; indeed, both
$\Delta U$ are consistent with the expected relation between the
ionsation parameters of the two different \glspl{sed}, namely $\Delta
\log U = \Delta \log Q = \log (Q_u Q_s^{-1})$\,$\simeq$\,$1.7$. A marginal
drop of the ionisation of the c--phase is also seen despite an
identical \gls{sed}, but we must stress that fixing the ionisation of
the s$\times$s$\times$u model to be the same as in the
u$\times$u$\times$u case, the fitting statistics is only marginally
worse ($\Delta C = 5$), so that an acceptable solution with the same
ionisation for the cold phase does exist.

According to the ionisation parameters of the three absorbers in the
s$\times$s$\times$u (Model\,3), one would assume that the l--phase
(with $\log U$\,$\simeq$\,$-1.0$) is now colder than the c--phase 
($\log U$\,$\simeq$\,$-0.6$). However, the two ionisation parameters can 
not be directly compared, as they are associated with two different
\glspl{sed}. In fact, the l--phase is still responsible for
higher--ionisation lines than the c--phase. This is shown in 
the right panels of 
Fig.\,\ref{fig:3phases_models} showing the best--fitting
{\small{PHASE}} models associated with the s$\times$s$\times$u
configuration (Model\,3). Comparing the right panels of 
Fig.\,\ref{fig:3phases_models} with its left panels 
(which are associated with the
u$\times$u$\times$u configuration, \ie\ with Model\,2), it is clear
that the best--fitting models for all three phases are only marginally
different, despite the h-- and l--phases have very different
ionisation parameters, and that the l--phase is always responsible for
higher--ionisation lines than the c--phase.

As a last step, in order to ease comparisons with other works\,/\,sources
when using different photoionisation models, we consider the {\small{XSTAR}}--based
{\small{ZXIPCF}} model in {\small{XSPEC}} and repeat the analysis by
using three {\small{ZXIPCF}} component instead of the three
{\small{PHASE}} ones. The free parameters of the
{\small{ZXIPCF}} model are the ionisation parameter, defined as $\xi =
L/(nr^2)$ where the luminosity is integrated between 1 and 1000\,Ry,
the column density, and the outflow velocity. As for
the turbulent velocity, it is fixed to $200$\,\kms\ in the
model. The intrinsic \gls{sed} is a slightly steeper power law of
$\Gamma = 2.2$ than for the {{\small{PHASE}} model we used so far
  ($\Gamma = 2.0$). 

The {\small{ZXIPCF}}--based model (Model\,4 in
Table\,\ref{tab:chandrafits}) is a significantly worse description of
  the data than the equivalent u$\times$u$\times$u model based on the
  {\small{PHASE}} code (Model\,2 in Table\,\ref{tab:chandrafits}). This
    is mostly due to a worse description of the ionised Fe absorption
    lines that are only partially reproduced by the model. Moreover
    the softest \xr\ energies (where Ne and Mg lines dominate) are
    not well reproduced, and residuals are also seen around 1.8\,\kev\ 
    (mostly Si lines). The best--fitting ionisation parameters are
    significantly higher than those of Model\,2. This is expected
    because of the different definition of ionisation parameters. In
    fact, assuming the same 2--10\,\kev\  luminosity for the two \glspl{sed}
    (one with $\Gamma=2$, and the other with $\Gamma=2.2$),
    the conversion between the {\small{ZXIPCF}} ionisation parameter
    and $U$ is $\log \xi = 1.99 + \log U$. Indeed, the ionisation
    parameters derived with the {\small{ZXIPCF}}--based Model\,4 and
    those derived via the {\small{PHASE}}--based Model\,2 are all
    consistent with that conversion within the errors (see
    Table\,\ref{tab:chandrafits}). Besides the different treatment of
    atomic physics, one further possible reason for the worse
    description of the data with Model\,4 with respect to Model\,2 is that
    the {\small{ZXIPCF}} model assumes a relatively modest turbulent
    velocity of $200$\,kms , while the best--fitting parameters
    in Model\,2 suggest a much higher v$_{\rm{turb}}$. Hence, we do not
    discuss any further the {\small{ZXIPCF}} model and we consider to
    have reached a satisfactory description of the \chandra\ 
    time--averaged spectrum with Model\,2 (or Model\,3).

The resulting picture is the following: during the $\sim$\,$10$~days
corresponding to the time--averaged \chandra\ spectrum, the
\xr\ continuum in \eso\ is transmitted through three absorbers, ionised at
different degrees. Although there is a slight difference in the
statistical results from the two physical scenarios defined above
(Model\,2 and Model\,3 in Table\,\ref{tab:chandrafits}), we can not claim
a highly significant preference for one of the two scenarios. For
simplicity, we adopt here the u$\times$u$\times$u Model\,2. This choice
allows us to compare directly the ionisation parameters of the
absorbers, as they all see the same \gls{sed}. Moreover, as discussed
above, there is no loss of generality because all parameters of
Model\,3 can be obtained from those of Model\,2 simply rescaling the
ionisation parameters taking into account the different photon rate
$Q$.

In this context, we characterise the warm absorbers (h-- and
l--phases) as two structures with ionisation parameters of $\log
U^{\rm{(h)}}$\,$\simeq$\,$1.9$ and $\log U^{\rm{(l)}}$\,$\simeq$\,$0.5$, 
and equivalent hydrogen column densities of the order of a few $\times
10^{23}$ and a few $\times 10^{22}$\,cm$^{-2}$ respectively. The third
absorber, previously thought to be neutral, turns out to be ionised,
with $\log U^{\rm{(c)}}$\,$\simeq$\,$-0.45$ and a column density of few
$\times 10^{22}$\,cm$^{-2}$. As per the outflow velocities, there are
no striking differences among the three ionisation phases, and the
most likely physical scenario appears to be one in which the warm
absorbers and the coldest one are all part of an outflow characterised
by velocities in the range of $\sim$\,$1000-2000$\,\kms , although the
l--phase appears to be marginally faster than the other two phases.

\section{Short time--scale absorbers variability}

Having reached a fair description of the time--averaged
\chandra\ data, we are now able to focus on the specific differences
among the four \chandra\ observations performed during ten days in
April 2010. We start by applying the best--fitting model described
above (Model\,2 in Table\,\ref{tab:chandrafits}) to the four
\chandra\ observations. The continuum photon index is initially free
to vary but, as no variability is seen after a few tests, it is forced
to be the same in all observations. On the other hand, the continuum
normalisation is free to vary independently.

\begin{table*} 
\caption{\label{tab:wabsvar}Best--fitting results for the analysis of
  the four 2010 \chandra\ observations from April 14 to April 24
  considered separately. The continuum photon index is initially free
  to vary but, after testing for its lack of variability, it is forced
  to be the same in all observations ($\Gamma = 1.97\pm 0.06$). In the
  upper part of the Table, we show the best--fitting results obtained
  if all the absorbers parameters are forced to be the same at all
  epochs (\ie\ we reproduce the results obtained using the
  time--averaged spectrum, see Table\,\ref{tab:chandrafits} for
  comparison). The middle part of the Table shows results obtained
  when the ionisation and column density of the three absorbing phases
  are free to vary independently in each of the four observations
  while keeping, for each phase, common outflow and turbulent
  velocities. Finally, the lower part of the Table shows results
  obtained by forcing all absorbers parameters to be the same at all
  epochs except the c--phase ionisation, which is the only one that
  varies significantly. As it was the case for the time--averaged
  \chandra\ spectra, the turbulent velocity of the h-- and l--phases
  is $\geq$\,$600$\,\kms\ (with a model upper limit of $900$\,\kms )
  while, in all fits, the c--phase is characterised by a turbulent
  velocity of $600\pm 200$\,\kms . Units are the same as in
  Table\,\ref{tab:chandrafits}. The symbol $p$ means that the
  parameter reached the limit allowed by the model. }
\begin{center}
\begin{tabular}{c c c c c c c c c c}
\hline\hline
\multicolumn{10}{c}{All phases forced to be constant at all epochs ($C/{\rm{dof}}=5351/4673$)} \\
\hline\hline
& \multicolumn{3}{c}{h--phase} &  \multicolumn{3}{c}{l--phase} & \multicolumn{3}{c}{c--phase} \\
\\
& $\log U$ & $\log {\rm{N_H}}$ & $v_{\rm outflow}$ & $\log U$ & $\log {\rm{N_H}}$ & $v_{\rm outflow}$ & $\log U$ & $\log {\rm{N_H}}$ & $v_{\rm outflow}$ \\
& $2.1\pm 0.3$ & $23.7^{+0.3p}_{-0.3}$ & $1200\pm 200$ & $0.5\pm 0.1$ & $22.40\pm 0.09$ & $1800\pm 300$ & $-0.44\pm 0.05$ & $22.70\pm 0.05$ & $1450\pm 250$ \\
\hline\hline
& \multicolumn{9}{c}{All phases allowed to vary in ionisation and column density ($C/{\rm{dof}}=5273/4655$)} \\
\hline\hline
&  \multicolumn{3}{c}{h--phase} &  \multicolumn{3}{c}{l--phase} & \multicolumn{3}{c}{c--phase} \\
\\
Obs. date & $\log U$ & $\log {\rm{N_H}}$ & $v_{\rm outflow}$ & $\log U$ & $\log {\rm{N_H}}$ & $v_{\rm outflow}$ & $\log U$ & $\log {\rm{N_H}}$ & $v_{\rm outflow}$ \\
\hline
Apr.~14 &$2.0\pm 0.4$ & $23.7^{+0.3p}_{-0.3}$ & $1250\pm 250$ & $0.5\pm 0.2$ & $22.2\pm 0.2$ & $1800\pm 350$ & $-0.30\pm 0.08$ & $22.7\pm 0.1$ & $1500\pm 250$ \\
Apr.~19 &$2.1\pm 0.4$ & $23.8^{+0.2p}_{-0.4}$ & ''  & $0.5\pm 0.2$ & $22.3\pm 0.2$ & ''  & $-0.60\pm 0.09$ & $22.7\pm 0.1$ & ''  \\
Apr.~21 &$2.1\pm 0.4$ & $23.8^{+0.2p}_{-0.4}$ & ''  & $0.5\pm 0.1$ & $22.4\pm 0.2$ & ''  & $-0.45\pm 0.08$ & $22.7\pm 0.1$ & ''  \\
Apr.~24 &$2.1\pm 0.4$ & $23.6^{+0.4p}_{-0.2}$ & ''  & $0.6\pm 0.1$ & $22.5\pm 0.2$ & ''  & $-0.34\pm 0.09$ & $22.7\pm 0.1$ & ''  \\
\hline\hline
& \multicolumn{9}{c}{All phases forced to be the same at all epochs, except the c--phase ionisation  ($C/{\rm{dof}}=5289/4670$)} \\
\hline\hline
&  \multicolumn{3}{c}{h--phase} &  \multicolumn{3}{c}{l--phase} & \multicolumn{3}{c}{c--phase} \\
\\
Obs. date & $\log U$ & $\log {\rm{N_H}}$ & $v_{\rm outflow}$ & $\log U$ & $\log {\rm{N_H}}$ & $v_{\rm outflow}$ & $\log U$ & $\log {\rm{N_H}}$ & $v_{\rm outflow}$ \\
\hline
Apr.~14 &$2.1\pm 0.3$ & $23.7^{+0.3p}_{-0.3}$ & $1200\pm 200$ & $0.5\pm 0.1$ & $22.40\pm 0.09$ & $1800\pm 300$ & $-0.31\pm 0.04$ & $22.70\pm 0.05$ & $1450\pm 250$ \\
Apr.~19 &''  & ''  & ''  & ''  & ''  & ''  & $-0.56\pm 0.07$ & ''  & ''  \\
Apr.~21 &''  & ''  & ''  & ''  & ''  & ''  & $-0.48\pm 0.06$ & ''  & ''  \\
Apr.~24 &''  & ''  & ''  & ''  & ''  & ''  & $-0.36\pm 0.04$ & ''  & ''  \\
\hline\hline
\end{tabular}
\end{center}
\end{table*}

We initially force all absorbers parameters to be the same in all
observations and we perform a joint ft to the four observations. This
is done to obtain a benchmark result that will be used to assess the
significance of any absorber variability once parameters will be let
free to vary independently in each observation. We reach a statistical
result of $C=5351$ for $4673$ \gls{dof}. All parameters are
consistent, within the errors, with those obtained from the
time--averaged spectrum (Table\,\ref{tab:chandrafits}) and they are
given in the upper part of Table\,\ref{tab:wabsvar}. 

In order to explore any absorber variability, we then leave the
ionisation and column density of the three phases free to vary
independently in the four observations, while keeping constant outflow
and turbulent velocities for each phase (this assumption turns out to
be justified by an a--posteriori check which shows no improvement if
these parameters are allowed to vary). The new best--fitting result is
$C=5273$ for 4655 \gls{dof}. The improvement is exclusively due to the
variability of the c--phase ionisation. All other parameters are
consistent with remaining constant in the four observations, as shown
in the middle part of Table\,\ref{tab:wabsvar}.

We then perform a final fit where all parameters are forced to be the
same in the four observations except the c--phase ionisation (\ie\ the
only parameter that is significantly variable). This is done to derive
the $\Delta C$ that is obtained when only the c--phase ionisation  is free
to vary. We obtain $C=5289$ for 4670 \gls{dof}, \ie\ an improvement by
$\Delta C = 62$ for 3 \gls{dof} with respect to the benchmark model (all
absorbers forced to be constant). This demonstrates that the
variability of the c--phase ionisation is highly significant, and that
a constant ionisation can be safely excluded. The best--fitting
parameters for this model are reported in the lower part of
Table\,\ref{tab:wabsvar}.

In order to gain some insights on the origin of the observed
variability, we consider here the relationship between the absorber
ionisation and the 2--10\,\kev\ nuclear luminosity. Under the
assumption of no intrinsic variation of the \gls{sed} shape between
the four \chandra\ observations (as it is the case, since the photon
index is consistent with being constant) and of the gas density and
location, the \xr\ luminosity is proportional to the photon rate $Q$
that enters the definition of ionisation parameter $U = Q(4\pi c n
R^2)^{-1}$, \ie\ doubling the luminosity implies that $Q$ and
therefore $U$ is twice as large (as the \gls{sed} is a simple power
law, the luminosity can be considered in any arbitrary band). Hence,
we do expect a linear relationship between ionisation $U$ and
luminosity if the absorbing gas phase is in photoionisation
equilibrium with the ionising continuum. We consider the
$U-L_{\rm{2-10}}$ relationship instead of the $U-Q$ one, because we
prefer to use two direct observables.

\begin{figure}
\begin{center}
\includegraphics[width=0.42\textwidth, angle=0]{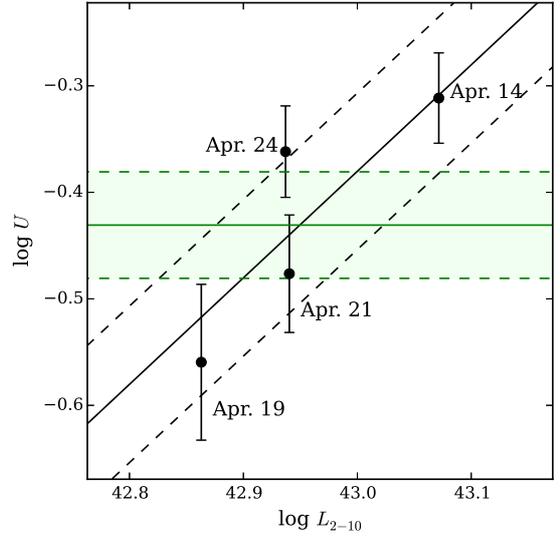}
\caption{\label{fig:u-lum} The ionisation of the c--phase component is
  shown as a function of the $2-10$\,\kev~\xr\ luminosity in
  $\log-\log$ space. The shaded area 
  between horizontal dashed lines is the best--fitting interval
  obtained when $U$ is forced to be the same at all epochs (see upper
  part of Table\,\ref{tab:wabsvar}). The data correspond to the case
  where all absorbers parameters, except the c--phase ionisation, are
  kept constant between the four \chandra\ observations (lower part of
  Table\,\ref{tab:wabsvar}). The tilted solid line is the
  best--fitting linear relation between $U$ and $L_{\rm{2-10}}$,
  \ie\ $\log U = a~\log L_{\rm{2-10}} +b$ with $a$ fixed to unity, and
  the tilted dashed lines represent the statistical fitting error given here
  at the 99 per cent confidence level. This shows that the data are
  consistent with a linear relation between $U$ and $L_{\rm{2-10}}$,
  as expected if the gas is in photoionisation equilibrium with the
  irradiating continuum. Letting $a$ free to vary confirms the linear
  relation although with relatively large error ($a=1.1\pm 0.4$). Note
  that some changes in gas properties (or intrinsic \gls{sed}) may be
  present between April 21 and 24, which would explain the slightly
  different ionisation of the c--phase despite an almost identical
  luminosity.}
\end{center}
\end{figure}

As can be seen in Table\,\ref{tab:wabsvar}, the ionisation of the h--
and l--phases is consistent with being constant during the 10~days
probed by the \chandra\ observations. However, the relatively large
errors imply that the two parameters could also be linearily related
to $L_{\rm{2-10}}$ at the same significance level, so that no
information can be obtained for the two warm\,/\,hot phases. Hence, we
focus here on the c--phase only, which is the only one showing
significant ionisation variability. Fig.\,\ref{fig:u-lum} shows the
c--phase ionisation as a function of the 2--10 \,\kev\ \xr\
luminosity (in $\log-\log$ space). The shaded area is the range of
$\log U$ obtained when all absorbers parameters (including the
c--phase ionisation) are forced to be the same at all epochs (see
upper part of Table\,\ref{tab:wabsvar}), while data points are those
reported in the lower part of the same Table. The solid line
represents the best--fitting relation of the form $\log U = a\log
L_{\rm{2-10}} + b$, where $a$ is fixed to unity to show that a linear
relationship between $U$ and $L_{\rm{2-10}}$ is indeed consistent with
the data. The tilted dashed lines represent the associated uncertainty (at
the 99 per cent confidence level). Letting $a$ free to vary confirms
the linear relation although with relatively large error ($a=1.1\pm
0.4$) and excludes, as expected, the case of constant $U$ ($a=0$).

We then conclude that the observed variability is driven by the linear
response of the c--phase ionisation to variations of the intrinsic
luminosity (or photon rate $Q$). This strongly suggests that the
c--phase is dense enough to be in photoionisation equilibrium with the
irradiating nuclear source on time--scales as short as a few days. As
mentioned, the h-- and l--phase are consistent with both constant
ionisation and with a linear relationship between $U$ and \xr\
luminosity, so that no information is gained in those cases. Repeating
the analysis with the s$\times$s$\times$u configuration (\ie\ assuming
Model\,3 of Table\,\ref{tab:chandrafits} as baseline instead of
Model\,2) gives very similar results, and only the c--phase ionisation
is significantly variable (and satisfies again a linear relationship
between $U$ and $L_{\rm{2-10}}$).

\section{The 2006 and 2013 \xmm\ observations}

The best--fit models reached in the previous Sections characterise the
ionisation states, equivalent column densities and outflow velocities
of three absorbers (all ionised at different degrees, and all with
similar outflow velocities), based on the rich set of absorption lines
detected in 2010 with \chandra. Here, we refine our results as far as
possible adding data from \xmm\ observations of 2006 and 2013 (pn),
that have been previously studied by \cite{bailon2008} and
\cite{miniutti2014} respectively.

The resolution of the \chandra\ \gls{hetgs} is much greater than that
of \xmm\ \gls{epic}-pn instrument, so that the pn spectra do not show
the profusion of lines that the HEG\,/\,MEG spectra do (as discussed
by \cite{miniutti2014} the \gls{rgs} data do not offer enough spectral
quality to be discussed in any of the available
\xmm\ observations). However, in order to try to assess the
variability of the aborbers detected with \chandra , we consider
here the \xmm\ pn spectra from the 2006 and 2013
observations. 

\begin{table*} 
\caption{\label{tab:2xmmcommon}Best--fitting common parameters for the
  simultaneous broad--band analysis of the 2006 and 2013
  \xmm\ observations. Luminosities are in units of
  $10^{42}$\,\ergps , temperatures and energies are given in
  \kev , the column density is in units of $10^{22}$\,cm$^{-2}$ and
  the line intensity is given in units of
  $10^{-6}$\,ph\,s$^{-1}$\,cm$^{-2}$. For parameters that are
  different between the two observations (nuclear continuum photon
  index and luminosity as well as its column density), see text.}
\begin{center}
\begin{tabular}{c c c c c }
\hline\hline
\multicolumn{5}{c}{Common constant components in the 2006\,/\,2013 \xmm\ observations} \\
\hline
Soft scatt. & \multicolumn{2}{c}{APEC (1)} & \multicolumn{2}{c}{APEC (2)} \\
$L_{\rm{0.5-2}}$ & $kT$ (keV) & $L_{\rm{0.5-2}}$ & $kT$ (keV) & $L_{\rm{0.5-2}}$ \\
$(6.7\pm 0.3)\times 10^{-2}$ & $0.80\pm 0.04$ & $(2.8\pm 0.2)\times 10^{-2}$& $0.09\pm 0.03$ & $(9\pm 1)\times 10^{-3}$\\
\hline
Cold refl. & \multicolumn{2}{c}{Hard scatt.} & \multicolumn{2}{c}{Ionised Fe line} \\
$L_{\rm{2-10}}$ & $L^{\rm scatt}$\,/\,$L^{\rm nucl}$ & N$_{\rm H}$ & $E_{\rm{rest}}$ & Intensity \\
$0.31\pm 0.02$ & $0.14\pm 0.03$ & $7.5\pm 0.8$ & $6.50\pm 0.04$ & $5.3\pm 1.7$ \\ 
\hline\hline
\end{tabular}
\end{center}
\end{table*}

The relatively simple baseline continuum model used so far is not an
adequate description of the \xmm\ data. This is for two main reasons:
i) the \xmm\ data extend down to 0.5\,\kev\  (as opposed to the
\chandra\ data that do not have sufficient signal--to--noise below
1.2\,\kev ) and reveal structure in the soft \xr s, and ii) the
\xmm\ observations (especially the 2013 one) are more heavily absorbed
than the \chandra\ one (see \eg\ Fig.\,\ref{fig:xmmchanxmm}) and an
additional hard \xr\ component is visible in the heavily absorbed
2013 \xmm\ spectrum.

As discussed extensively by \cite{miniutti2014}, the soft \xr\
structure is well reproduced by adding two thermal plasma components
(the {\small{APEC}} model in {\small{XSPEC}}) to the soft power law
which represents the scattered re--emission of the nuclear continuum
by some extended gas. On the other hand, the 2013 \xmm\ spectrum
requires an additional power law in the hard \xr\ band, absorbed by
a column density that is different from that affecting the nuclear
emission. This component was interpreted by \cite{miniutti2014} as
scattered emission in a clumpy absorber. The idea is that, if the main
absorber is clumpy instead of homogeneous, the observed spectrum
should comprise not only a transmitted component (\ie\ the nuclear
continuum absorbed by the particular clump that happens to be in our
\gls{los}) but also a scattered component due to clumps out of the
\gls{los} that intercept the nuclear emission re--directing part of it
into the \gls{los}. As this scattered component reaches us after
passing through the clumpy absorber itself, it is absorbed by the
spatially--averaged column density of the clumps rather than by that
of the particular one that is in the \gls{los} at the given
epoch. This explains why the scattered component is absorbed by a
different column density than the nuclear continuum, although the
scattered fraction (\ie\ basically the ratio between the scattered and
nuclear luminosity) and the column density towards the scattered
component should be the same at all epochs (as they represent
spatially--averaged values). \cite{miniutti2014} have derived a
scattered fraction of $15\pm 3$~per cent and a (neutral) column
density of $(7.6\pm 0.8)\times 10^{22}$\,cm$^{-2}$ towards the
scattered component. Note that the scattered component makes a
non--negligible contribution to the hard \xr\ spectrum only when the
nuclear continuum is sufficiently absorbed (\ie\ for column densities
significantly higher than $7\times 10^{22}$\,cm$^{-2}$), which only
occurs during the 2013 \xmm\ observation.

We then consider a new baseline continuum model comprising all the
above components. The global baseline model and its main 
components are shown in Fig.\,3 of \cite{miniutti2014}. We
first perform a joint fit to the two \xmm\ observations and we force
all parameters that are not expected to vary between the two epochs to
be the same, namely i) the temperature and normalisation of the two
{\small{APEC}} components, ii) the photon index and normalisation of
the soft scattered power law, iii) the column density towards the hard
scattered power law as well as the hard scattered fraction, and iv)
the flux of the neutral reflection component (carrying the \fe\ emission line). The best--fitting model produces $\chi^2=1649$
for 1445 \gls{dof}. Close inspection of the \fe\ region reveals the
presence of some emission structure bluewards of the line, especially
in the more heavily absorbed 2013 \xmm\ observation. We add a Gaussian
emission line forcing it to have the same properties in the two
observations and we reach $\chi^2=1612$ for 1443 \gls{dof} for an additional
Gaussian emission line at $6.50\pm 0.04$\,\kev\  (in the galaxy
rest--frame). When the line intensity is allowed to vary between the
two observations, no improvement is obtained, and the line \gls{ew} 
is $\sim$\,$35$\,\ev\ in 2006 and $\sim$\,$85$\,\ev\ in 2013, reflecting the
lower \xr\ continuum in 2013. If no energy shift is assumed, the
line most likely arises from ionised Fe emission.

All parameters that are kept in common between the two observations
are reported in Table\,\ref{tab:2xmmcommon}. As for the variable
parameters, they are the nuclear continuum photon index and
normalisation, and the (neutral) column density towards it. As already
shown by \cite{miniutti2014}, the 2006 \xmm\ observation is absorbed by
a column of $\sim$\,$5.6\times 10^{22}$\,cm$^{-2}$, while the column
density during the 2013 observation is one order of magnitude
higher. The two photon indices are consistent with each other within
the errors ($\Gamma = 1.95\pm 0.07$ in 2006 and $2.0\pm 0.1$ in 2013).

Having derived a baseline continuum model for the two observations, we
now search for signatures of absorption features. All common parameters are fixed to their
best--fitting values obtained from the joint fit of the two
observations (see Table\,\ref{tab:2xmmcommon}) and we consider the two
observations separately. The baseline model (with common parameters
fixed at their best--fitting values) results in  $\chi^2=829$ for 705 \gls{dof}
for the 2006 observation and $\chi^2=783$ for 749 \gls{dof} for the 2013
one. We first add a series of Gaussian absorption lines with width
fixed at $1$\,\ev\ and redshift fixed at the galaxy one, and we report
the lines improving the fit by more than $\Delta \chi^2 = 9.2$ in
Table\,\ref{tab:2xmmlines} together with their possible
identification for the 2006 and 2013 observations. 

\begin{table*} 
\caption{\label{tab:2xmmlines}Absorption lines detected with Gaussian
  models in the \xmm\ \gls{epic}-pn spectra. Every line contributes
  with two free parameters (rest--frame energy and intensity). Only
  lines producing a statistical improvement of 
  $\Delta \chi^2$\,$\geq$\,$9.2$ are reported. }
\begin{center}
\begin{tabular}{l l c c c c c c}
\hline\hline 
\multicolumn{8}{c}{2006 \xmm\ observation} \\
\hline
Phase & ID & Transition & E$_{\rm lab}$\,(\kev)\,/\,$\lambda_{\rm lab}$\,(\r{A}) & E$_{\rm restframe}$\,(\kev)& $-$EW\,(eV)& v$_{\rm outflow}$\,(\kms) & $\Delta \chi^2$ \\
\hline
 h & S\,\textsc{xvi}   & $1s   \rightarrow 2p$    & $2.6218$\,/\,$4.729$ & $2.64\pm 0.03$ & $20\pm 10$ & $2050\pm 3350$ & $10$ \\
 h & Fe\,\textsc{xxv}  & $1s^2 \rightarrow 1s 2p$ & $6.7019$\,/\,$1.850$ & $6.75\pm 0.03$ & $80\pm 20$ & $2150\pm 1300$ & $44$ \\
 h & Fe\,\textsc{xxvi} & $1s   \rightarrow 2p$    & $6.9650$\,/\,$1.780$ & $7.06\pm 0.04$ & $90\pm 20$ & $4050\pm 1350$ & $37$ \\
$-$  & $-$ & $-$ & $-$ & $7.7\pm 0.1$ & $60\pm 30$ & $-$ & $11$ \\
 $-$ & $-$ & $-$ & $-$ & $8.4\pm 0.1$ & $60\pm 30$ & $-$ & $11$ \\
\hline\hline
\multicolumn{8}{c}{2013 \xmm\ observation} \\
\hline
Phase & ID & Transition & E$_{\rm lab}$\,(\kev)\,/\,$\lambda_{\rm lab}$\,(\r{A}) & E$_{\rm restframe}$\,(\kev)& $-$EW\,(eV)& v$_{\rm outflow}$\,(\kms) & $\Delta \chi^2$ \\
\hline 
 $-$ & $-$ & $-$ & $-$ & $2.15\pm 0.03$ & $40\pm 15$ & $-$ & $18$ \\
\hline\hline
\end{tabular}
\end{center}
\end{table*}

\subsection{The 2006 \xmm\ observation}

The two more significant lines are detected at $6.75\pm 0.03$\,\kev\ and
at $7.06\pm 0.04$\,\kev .  If identified with Fe\,\textsc{xxv} and
Fe\,\textsc{xxvi} respectively, the observed lines are associated with
high outflow velocities and, considering the large errors,
with a possible common outflow velocity of $3100\pm400$\,\kms . 
This outflow velocity is significanlty higher than
that derived from the \chandra\ data from the same two absorption
lines (which have a common outflow velocity of $1150\pm
350$\,\kms ). The absorption line at $\sim$\,$2.64$\,\kev\  (see
Table\,\ref{tab:2xmmlines}) is not highly significant, and can
be probably associated with S\,\textsc{xvi}, although the very large
error on the line energy does not allow to place a secure label on
this feature. If the line is indeed associated with
S\,\textsc{xvi}, the outflow velocity is consistent with that of the
two ionised Fe lines (although with large error), possibly indicating
a common origin in a highly--ionised wind. The other two, less
significant absorption lines in the 2006 observation have no clear
identification (see Table\,\ref{tab:2xmmlines}).

The detection of Fe\,\textsc{xxv} and Fe\,\textsc{xxvi} absorption
lines means that a hot phase with similar ionisation as that detected
in 2010 with \chandra\ is present in the 2006 observation as well, so
we replace all Gaussian absorption lines with a {\small{PHASE}}
model. The {\small{PHASE}} model produces a significant improvement,
and we reach $\chi^2 = 712$ for 701 \gls{dof} (to be compared with
$\chi^2=829$ for 705 \gls{dof} of the baseline continuum model). The
absorber has $\log U = 2.0\pm 0.3$, a column density of $\log
{\rm{N_{H}}}$\,$\geq$\,$23.4$, and an outflow velocity of $2700\pm
1100$\,\kms , consistent within the relatively large error with the
common outflow velocity of $3100\pm400$\,\kms\ derived from the energy
shift of the Fe\,\textsc{xxv} and Fe\,\textsc{xxvi} absorption lines.

As can be seen in Table\,\ref{tab:2xmmlines}, we do not find any
absorption line that may be related to the l-- or c--phases detected
in the high--resolution \chandra\ data. However, it is interesting to
see if gas with similar properties as seen in the 2010
\chandra\ observation is consistent with the 2006 \xmm\ data as
well. We first add an l--phase, fixing all of its parameters to those
detected with \chandra\ (see Table\,\ref{tab:chandrafits},
Model\,2). The statistical result ($\chi^2 = 710$ for 701 \gls{dof})
can not be distinguished from the one with no l--phase ($\chi^2 = 712$
for the same number of \gls{dof}). Letting all the parameters of the
l--phase free to vary does not produce any improvement. We conclude
that gas with the same properties as the l--phase detected with
\chandra\ in 2010 is neither required nor excluded by the \xmm\ data.
Its non--detection is likely due to i) the worse energy--resolution of
the pn data, and ii) the higher column density of the coldest phase
which lowers significantly the signal--to--noise ratio in the relevant
spectral region (starting to be dominated by extended emission and
scattered light rather than by the \xr\ nuclear continuum).

As for the strictly neutral absorber (with column density of 
$\sim$\,$5.6\times 10^{22}$\,cm$^{-2}$), we replace it with an ionised
{\small{PHASE}} model. As no absorption lines associated with that
component are detected, we fix its outflow velocity to that detected
with \chandra , namely $1400$\,\kms\ (see Model\,2 in
Table\,\ref{tab:chandrafits}). The replacement, however, does not
produce any improvement, and the resulting ionisation parameter is
only an upper limit of $\log U$\,$\leqsim$\,$-0.3$. Hence, a non--zero
ionisation of the coldest phase is neither required nor excluded by
the \xmm\ data. The column density of this coldest phase presents some
degree of degeneracy with the (basically unconstrained) ionisation
parameter and can take any value between $\log {\rm{N_{H}}}$\,$\simeq$\,$22.7$ 
and $\log {\rm{N_{H}}}$\,$\simeq$\,$23.3$ (the lower column density
being associated with the lowest possible ionisation, $\log U = -3$). 

However, as shown in the previous Section, the c--phase 
ionisation responds to the continuum variations on short 
time--scales during the 2010 \chandra\ 10~days--long monitoring, and 
this response can be used to break the degeneracy between 
ionisation and column density in the 2006 data. This is because, 
assuming a perfectly homogeneous c--phase even on long time--scales 
(\ie\ no clumpiness for the c--phase) the 2006 ionisation can be 
predicted from the 2010 one by considering the different intrinsic 
continuum luminosity at the two epochs. For a fixed \gls{sed} 
shape (and the photon indices in 2006 and 2010 are consistent with 
each other), the 2--10\,\kev\ \xr\ intrinsic luminosity is 
proportional to the photon rate $Q$ and hence to the ionisation 
state $U$. As the c--phase ionisation during the 2010 
\chandra\ observation is $\log U = -0.45$ (see Model\,2 in 
Table\,\ref{tab:chandrafits}) and since the intrinsic 2--10\,\kev\ 
luminosity in 2006 is a factor of $\sim$\,$1.3$ lower than in 2010, 
the 2006 c--phase must have ionisation $\log U = -0.56$ if the 
c--phase is homogeneous (as opposed to clumpy) on long time--scales. 

This removes the degeneracy between column density and ionisation 
during the 2006 \xmm\ observation, and allows us to check whether the 
absorber is indeed homogeneous as opposed to clumpy, by comparing the 
2006 and 2010 column densities directly. This comparison is shown in 
Fig.\,\ref{fig:contNH} in terms of the statistical improvement as a 
function of column density for the 2010 (left) and 2006 (right) 
observations. Although the difference is small in absolute terms, the 
two column densities are inconsistent with each other at more than 
$5\sigma$, indicating the non--homogeneous nature of the c--phase 
absorber on long time--scales. A clumpy absorber is also strongly 
suggested by the presence of a hard scattered component during the 
much more heavily absorbed 2013 \xmm\ observation as discussed in 
detail by \cite{miniutti2014}. 

\begin{figure}
\begin{center}
\includegraphics[height=0.42\textwidth,angle=-90]{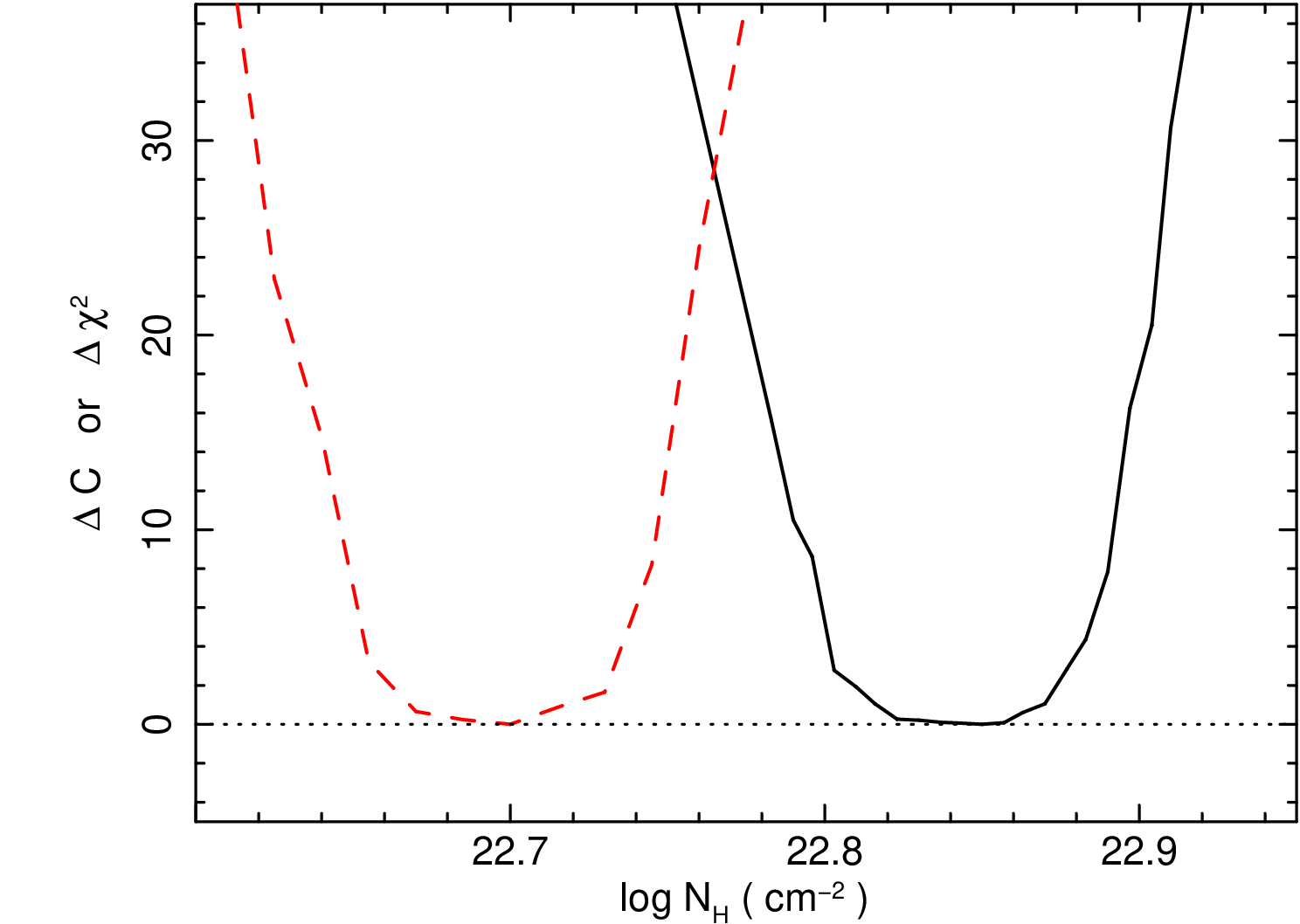}
\caption{\label{fig:contNH} The statistical improvement as a function
  of the c--phase column density for the 2010 \chandra\ (left) and the
  2006 \xmm\ (right) observations as obtained with the
  {\small{STEPPAR}} command in {\small{XSPEC}}. We use the $\Delta C$
  and $\Delta \chi^2$ values for the \chandra\ and the \xmm\ data
  respectively. All parameters are free to vary during the
  minimisation except the ionisation of the c--phase during the 2006
  \xmm\ observation. This ionisation is fixed to that expected under
  the assumption of a homogeneous rather than clumpy absorber ($\log
  U = -0.56$, see text for details). The column densities of the
  c--phase at the two epochs are similar but inconsistent with each
  other at more than 5$\sigma$, strongly suggesting the inhomogeneous
  nature of the c--phase.}
\end{center}
\end{figure}

We do not replace the
neutral absorber towards the hard scattered component with an ionised
one because this component has an almost negligible contribution in
the 2006 \xmm\ observation and it is only strictly required by the
2013 \xmm\ data \citep{miniutti2014}.

The best--fitting parameters for our final model, comprising the
highly ionised h--phases and the two neutral absorbers towards the
continuum and the hard scattered component are reported in
Table\,\ref{tab:longterm}. Fig.\,\ref{fig:xmmlines} (top) shows the
high--energy spectrum from the 2006 observation together with our
best--fitting model, dominated by the Fe emission lines at
$6.4$\,\kev\ and $\sim$\,$6.5$\,\kev , and by the h--phase component.

\begin{table} 
\caption{\label{tab:longterm}Best--fitting parameters for the
  \xmm\ observations. For the 2006 observation, the two c--phases are
  here modelled with a strictly neutral absorber at rest, as no
  improvement is obtained by replacing them with a {\small{PHASE}}
  model (see text for details on the parameters obtained when an
  ionised model is used instead). The overscript $^f$ means that the
  parameter is fixed, the symbol $p$ indicates that the error on one
  parameter reached the limit of the model. Units as in
  Table\,\ref{tab:chandrafits}.}
\begin{center}
\begin{tabular}{c c c c c}
\hline\hline 
& & & $2006$ & $2013$ \\ 
\hline 
& & $\Gamma$ & $1.95 \pm 0.07$ & $2.0\pm 0.1$ \\ 
& & $L^{\rm{nucl}}_{2-10}$ & $7.5$ & $5.8$ \\ 
\hline
\multirow{4}{*}{\begin{turn}{90}Warm abs.\end{turn}} &
\multirow{4}{*}{\begin{turn}{90}h--phase\end{turn}} 
& log\,${\rm U}$ & $2.0\pm 0.3$ & $2.0\pm 0.3$ \\
& & log\,N$_{\rm H}$ & $23.7^{+0.3p}_{-0.3}$ & $23.6^{+0.4p}_{-0.4}$ \\ 
& & v$_{\rm turb}$ & $100-900$ & $100-900$ \\ 
& & v$_{\rm outflow}$ & $2700\pm 1100$ & $4100\pm 1700$ \\ 
\hline \multirow{4}{*}{\begin{turn}{90}Cold abs.\end{turn}} &
\multirow{4}{*}{\begin{turn}{90}c--phase\end{turn}} 
& log\,${\rm U}$ & $-$ & $0.2\pm 0.2$ \\ 
& & log\,N$_{\rm H}$ & $22.7\pm 0.2$ & $23.8\pm 0.1$ \\ 
& & v$_{\rm turb}$ & $-$ & $100-900$ \\ 
& & v$_{\rm outflow}$ & $0^f$ & $1400^f$ \\
\hline 
\multirow{4}{*}{\begin{turn}{90}Scatt. abs.\end{turn}} & 
\multirow{4}{*}{\begin{turn}{90}c--phase\end{turn}} 
& log\,${\rm U}$ & $-$ & $\leqsim$\,$-0.3$ \\
& & log\,N$_{\rm H}$ & $22.9^f$ & $22.9\pm 0.2$\\ 
& & v$_{\rm turb}$ & $-$ & $100-900$ \\ 
& & v$_{\rm outflow}$ & $0^f$ & $1400^f$ \\ 
\hline 
& & $\chi^2$\,/\,dof &$712$\,/\,$701$ & $754$\,/\,$739$ \\ 
\hline\hline
\end{tabular}
\end{center}
\end{table}

\subsection{The 2013 \xmm\ observation}

The only significant absorption line detected in the 2013
\xmm\ spectrum is at $\sim$\,$2.15$\,\kev\  (see
Table\,\ref{tab:2xmmlines}). Its identification is difficult,
and we also note that the feature occurs close to where a significant
drop in quantum efficiency is seen for the pn detector, so that we
can not exclude that it has an instrumental origin. Hints for
high--energy absorption lines bluewards of the \fe\ emission
line are seen, but none reaches the required $\Delta \chi^2 = 9.2$ when
Gaussian models are considered.

With no absorption lines, applying detailed photoionisation codes to
the 2013 data will likely result in over--modelling. However, the
nuclear and scattered continua are still absorbed by strictly neutral
matter, while from the \chandra\ observations, we have indications that even the coldest component is
in fact ionised. Note also that, based on the long--term variability
properties of the coldest absorber, \cite{miniutti2014} have
interpreted the 2013 data with absorption from a cloud of the \gls{blr}
as opposed to the 2006 \xmm\ and 2010 \chandra\ observations where the
c--phase was associated with a clumpy absorber at the torus scale.

\begin{figure}
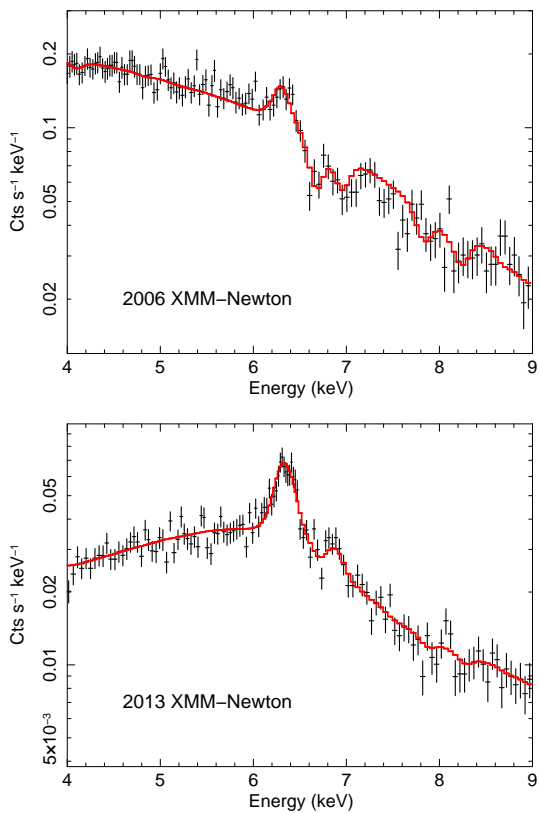

\begin{center}

\includegraphics[height=0.42\textwidth,angle=-90]{xmmold_phasesall.eps}

\vspace{0.3cm}

\includegraphics[height=0.42\textwidth,angle=-90]{xmmnew_phasesall.eps}
\caption{\label{fig:xmmlines} The high--energy part of the
  \xmm\ \gls{epic}--pn spectra from the 2006 (top) and 2013 (bottom)
  observations is shown together with the best--fitting model. The
  high--energy absorption features are well reproduced by the
  h--phase, although the significance of this component is realtively
  low in the 2013 data. The improvement of the statistical result with
  the addition of the h--phase is $\Delta \chi^2 = 117$ for 4 \gls{dof} in the
  2006 data and only $\Delta \chi^2 = 14$ in the 2013 data. Data have been
  slightly re--binned for visual clarity.}
\end{center}
\end{figure}

We test this scenario in the following way: we first replace the
neutral absorber towards the nuclear continuum with a {\small{PHASE}}
model. The lack of absorption features means that we are unlikely to
be sensitive to outflow velocities, and we choose to fix it to the
2010 c--phase outflow velocity ($1400$\,\kms ). The model produces
$\chi^2 = 772$ for 747 \gls{dof}, slightly better than the baseline
($\chi^2 = 783$ for 749 \gls{dof}). We obtain $\log U^{\rm{nucl}}$\,$\sim$\,$0.2$ 
with $\log {\rm{N^{nucl}_{H}}}$\,$\sim$\,$23.8$ for the c--phase
towards the nuclear continuum. Hence, the main absorber in 2013 is
significantly more highly ionised than that in 2006 and 2010 (as well
as being associated with a much higher column density). As for the
hard scattered component absorber, replacing it with a {\small{PHASE}}
model outflowing at $1400$\,\kms\ does provide a very marginal
improvement ($\chi^2 = 768$ for 743 \gls{dof}), but its ionisation is
only an upper limit of $\log U^{\rm{scatt}}$\,$\leqsim$\,$-0.3$. Letting the
outflow velocities of the two absorbers free to vary does not provide
any improvement, and the large errors are consistent with zero
velocity (as well as with velocities of a few thousands \kms\ in both
outflow or inflow), showing that we are not sensitive to this
parameter, as expected.

As some hints for high--energy absorption lines are seen in the data,
we add a highly ionised {\small{PHASE}} component to assess whether
the h--phase can be detected in the 2013 \xmm\ data. In this case, we
leave the outflow velocity free to vary, as it may be possible to
constrain it if a set of low \gls{ew} absorption features is present
in the data (although none is individually significant). We reach a
best--fit of $\chi^2 = 754$ for 739 \gls{dof}, marginally better than
that with no h--phase component ($\chi^2 = 768$ for 743
\gls{dof}). The h--phase has $\log U$\,$\sim$\,$2.0$, 
$\log {\rm{N_{H}}}$\,$\sim$\,$23.6$, and an outflow velocity of 
$4100\pm 1700$\,\kms . As it was the
case for the 2006 \xmm\ observation adding an l--phase with the same
properties as that detected in 2010 with \chandra\ does not improve
nor worsen the best--fit. Hence such a component is not rquired by the
data, but its presence can not be excluded either. Our final
best--fitting parameters are reported in Table\,\ref{tab:longterm} and
the high--energy spectrum from the 2013 observation is shown in the
bottom panel of Fig.\,\ref{fig:xmmlines}, where the
(low--significance) h--phase imprints some weak absorption features.

Summarising, the 2006 observation shows a clear h--phase with
ionisation and column density of the order of those detected with the
high--resolution \chandra\ data (note, however, that its outflow
velocity is significantly higher). Although the 2006 data do not allow
us to distinguish between a strictly neutral and an ionised c--phase,
we have shown that the 2006 c--phase column density is most likely different
from the 2010 one, strongly suggesting that the c--phase is not
homogeneous on long time--scales, but rather most likely clumpy. 

In 2013, the nuclear continuum is absorbed by a
much higher column (close to $10^{24}$\,cm$^{-2}$) of higher
ionisation gas ($\log U$\,$\sim$\,$0.2$), and a low--significance h--phase
is detected with similar properties to that observed in 2006. The
heavily absorbed 2013 \xr\ spectrum reveals the presence of a hard
scattered component. The scattered fraction is $\sim$\,$14$~per cent, and
the scattered emission is transmitted through a gas phase with 
$\log U$\,$\leqsim$\,$-0.3$ and $\log {\rm{N_H}} \sim 22.9$. A phase 
with intermediate ionisation is not required by the 2006 and 2013
\xmm\ data, but an l--phase with the same properties as that detected
with \chandra\ in 2010 is consistent with both data sets.

\section{Discussion}
\label{sec:discussion}

The temperature and ionisation state of the three absorbers we have 
detected with \chandra\ are similar to those of warm absorbers typically
detected in Seyfert\,1 galaxies. Column densities are however on the
high side for standard warm absorbers, see \eg\ \cite{Torresi10} or
\cite{Gupta13}. Moreover, the column density of the coldest absorbing
system is of the same order, or even higher, than that of the higher
ionisation components while, in general, higher ionisations are
associated with larger columns, \eg\ \cite{zhang15}. These differences
can be interpreted naturally within a scenario in which
\eso\ represents a source intermediate between Seyfert\,1 and Seyfert\,2
galaxies. In fact, considering the polar--scattered nature of the
source, and its likely intermediate inclination of
$\sim$\,$45^\circ$ \citep{schmid2003}, \eso\ appears to lie precisely in
that intermediate--inclination region of the parameter space.  In this
framework, the cold absorber we detect would be the same warm absorber
as seen in other sources, with its special properties due to the
orientation at which we are seeing this source: grazing the edge of
the clumpy torus. Objects observed at higher inclination angles (with
respect to the symmetry axis) than \eso\ correspond to highly \xr\ 
obscured Seyfert\,2 galaxies; sources viewed at slightly lower
inclination angles are typical Seyfert\,1 galaxies with standard warm
absorbers; objects at even lower inclination angles result into
a Seyfert\,1 optical classification and are likely to be characterised
by the absence of warm absorbers (\ie\ polar Seyfert\,1 galaxies
offering a naked view of the innermost \xr\ emitting region).

The three absorbers we detect in the high--resolution \chandra\ data
have similar outflow velocities, suggesting a common origin. Only the
coldest one responds with ionisation changes to intrinsic luminosity
variations on short time--scales (days). The coldest absorber is most
likely clumpy, as suggested by the comparison between the column
densities derived in 2010 and in 2006, and by the need for an extra
hard scattered component in the heavily absorbed 2013 observation, as
discussed already by \cite{miniutti2014}. On the other hand, the
highest--ionisation phase is consistent with the same ionisation and
column density on both short and long time--scales, although its
outflow velocity appears to be variable on long time--scales. As for
the intermediate zone (the l--phase), it is only detected in the
high--resolution, relatively unobscured
\chandra\ observation. However, a phase with similar properties is
allowed (not required) to be present in the \xmm\ data as well. This
suggests that the warm absorbers are distributed in a more homogeneous
way than the coldest one. Such properties (similar outflow velocities,
clumpiness of the denser, colder absorber, and relatively homogeneous
distribution of the warmer phases) suggest to consider a solution in
which all absorbers are part of the same outflow, with the warmer
phases providing the pressure confinement that is necessary to support
relatively long--lived, dense, colder clouds (although originally
considered for the \gls{blr} case, see \eg\ \cite{krolik81} for the
multi--phase confinement model, and \cite{emmering92} and
\cite{elvis2000} for the idea of an outflowing multi--phase structure
of the \gls{blr}).

Such a scenario can not be directly confirmed by the data, but we can
at least test whether the three absorbers detected in the 2010
\chandra\ observation are consistent with being in pressure
equilibrium. In order to derive the properties of the three gas
phases, we need to estimate the product $nR^2$ which, by definition, only depend
on the ratio between photon rate and ionisation ($Q/U$). This means
that deriving the gas properties from the u$\times$u$\times$u
configuration (Model\,2 in Table\,\ref{tab:chandrafits}) or from the
s$\times$s$\times$u one (Model\,3) gives the same result (we have shown
in Section\,\ref{sec:global} that, for the h-- and l--phases, the drop
in $U$ obtained with Model\,3 is simply proportional to the drop in $Q$
due to the different, absorbed \gls{sed}). For simplicity, as well as
for consistency with the previous Sections, we continue to use the
u$\times$u$\times$u configuration (Model\,2) as best--fitting model
from which we derive estimates on the gas properties.- As mentioned,
consistent results are obtained using Model\,3 instead.

\subsection{The ``cold'' absorber in 2010}

We proceed with orders--of--magnitude estimates of the properties of
the absorbing gas. According to our best--fitting model (Model\,2 in
Table\,\ref{tab:chandrafits}), the photon rate integrated from 1\,Ry to
infinity is $Q$\,$\simeq$\,$1.8\times 10^{53}$\,ph\,s$^{-1}$ during the
10~days probed by the \chandra\ time--averaged spectrum.

From the fact that the cold absorber is in photoionisation equilibrium
with the impinging continuum within five days, as shown in 
Fig.\,\ref{fig:u-lum}, and using the formula and procedures of
\cite{nicastro99} with the recombination time--scales and fractions
for Si\,\textsc{viii--x} for the best--fitting ionisation parameter
log\,${\rm U^{\rm{(c)}}}$\,=\,$-0.45$, one can estimate a lower limit
on the cold absorber number density of 
$n^{\rm{(c)}}$\,$\geqsim$\,$5\times10^4$\,cm$^{-3}$ 
(using other ions produces a difference of a
few per cent only). Using the definition of $U$, one has that 
$R^{\rm{(c)}}$\,$\leqsim$\,$5\times 10^{18}$\,cm\,$\simeq$\,$1.6$\,pc. 
A lower limit on the location of 
the c--phase can be obtained by considering that the observed outflow 
velocity ($\sim$\,$1400$\,kms) can not be lower than the escape 
velocity at the minimum launching radius. Knowing that the black hole 
mass is $M_{\rm{BH}}$\,$\simeq$\,$2.5\times 10^7$\,$M_\odot$ \citep{wang07}, we 
derive $R^{\rm{(c)}}$\,$\geqsim$\,$3 \times 10^{17}$\,cm\,$\simeq$\,$0.1$\,pc. Note
that the dust sublimation radius in \eso\ is located at
$\sim$\,$0.14$\,pc \citep{miniutti2014}, so that the c--phase is
consistent with being located somewhere between the inner and outer
edges of the so--called torus \citep{krolik2001}.  This conclusion agrees well with
that reached by \eg\ \cite{Blustin05} who, based on the analysis of
high energy--resolution data on a sample of 23 \gls{agn}, have shown that
warm absorbers most likely originate as outflows from the dusty torus.

Using again the definition of ionisation parameter, the lower limit on
$R^{\rm{(c)}}$ translates into an upper limit on the number density,
namely $n^{\rm{(c)}}$\,$\leqsim$\,$10^7$\,cm$^{-3}$. Finally, using
derived values of $n^{\rm{(c)}}$, the observed column density $\log
{\rm{N_H^{(c)}}}$\,$\simeq$\,$22.7$, and assuming a spherical absorbing
cloud, the diameter of the cloud $D^{\rm{(c)}}$ is in the range of
[$5\times10^{15}-10^{18}$]\,cm. Having constrained the absorber
density and knowing its equilibrium temperature (which is 
$\sim$\,$5\times 10^4$\,K according to our best--fitting {\small{PHASE}} model),
the pressure of this component can be estimated as $P = k_{\rm B} n T$
(where $k_{\rm B}$ is the Boltzmann constant), so that 
$P^{\rm{(c)}} = [3\times10^{-7} - 7\times 10^{-5}]$\,dyne\,cm$^{-2}$. 

\subsection{The two warm absorbers in 2010}

In order to check if the warm\,/\,hot absorbers can
pressure--confine the clumpy coldest phase, we assume here
co--spatiality of the three phases, as suggested by the similar
outflow velocities. Hence, for all absorbers, we assume $R =
[0.3-5]\times 10^{18}$\,cm. Using $Q$\,$\simeq$\,$1.8\times
10^{53}$\,ph\,s$^{-1}$ and the best--fitting ionisation parameters of
the h-- and l--phases ($\log U^{\rm{(h)}}$\,$\simeq$\,$1.9$ and 
$\log U^{\rm{(h)}}$\,$\simeq$\,$0.5$), one can constrain the number 
densities of the two warm absorbers to be 
$n^{\rm{(h)}} = [2\times10^2 - 7\times10^4]$\,cm$^{-3}$ 
and 
$n^{\rm{(l)}} = [6\times10^3 - 2\times10^6]$\,cm$^{-3}$ 
respectively. According to our best--fitting model,
the temperature of the two warm absorbers is $5.8\times 10^6$\,K for
the h--phase and $5.5\times 10^5$\,K for the l--phase, so that their
pressure is 
$P^{\rm{(h)}} = [10^{-7} - 6\times10^{-5}]$\,dyne\,cm$^{-2}$ 
and 
$P^{\rm{(l)}} = [4\times10^{-7} - 10^{-4}]$\,dyne\,cm$^{-2}$ 
respectively. The pressure of the warm
absorbers is in both cases consistent with that derived for the
c--phase. The common pressure interval for the three phases to be in
pressure equilibrium is 
$P^{\rm{(c,h,l)}} = [4\times10^{-7} - 6\times 10^{-5}]$\,dyne\,cm$^{-2}$.

We then conclude that the wind we detect in the data from the 2010
\chandra\ observations is consistent with being associated with the
atmosphere of the clumpy torus, and with being arranged in three
phases in pressure equilibrium with each other, with the coldest
clumps being confined by the more homogeneous and hotter phases. Note
that we can not claim that both the h-- and l--phases contribute to the
confinement, but only that they are both consistent with this role
which may be well dominated by one of the two phases.

\subsection{The absorbers in 2006 and 2013}
\label{subsec:abs0613}

The ionisation of the c--phase is only poorly constrained during the
2006 \xmm\ observation. However, its column density is clearly higher
than that during the 2010 \chandra\ observation, demonstrating the
clumpiness of this component. The truly different case is represented by the
2013 \xmm\ observation, where a much higher column density and
ionisation for the c--phase are obtained. As mentioned,
\cite{miniutti2014} have interpreted the c--phase in 2013 as due to a
clump (cloud) of the \gls{blr}, as opposed to the 2010 (and
probably 2006) observations where we just show that the c--phase is
most likely associated with the clumpy torus.

We first assume that the 2013 c--phase has the same origin as in 2006
and 2010, namely that it is confined within the inner and outer edges
of the torus. According to our best--fitting model, the photon rate in
the 2013 \xmm\ observation is $Q$\,$\simeq$\,$10^{53}$\,ph\,s$^{-1}$
and the best--fitting ionisation of the cold phase is $\log U$\,$\simeq$\,$0.2$. 
Hence, if the 2013 absorber was co--spatial with the c--phase
detected in 2010 with \chandra, its density should be $\sim$\,$8$ times
lower than in 2010 ($(Q/U)^{\rm{(2013)}}$\,$\simeq$\,$(Q/U)^{\rm{(2010)}}/8$). 
Given the much higher column density observed
in 2013 (${\rm{N_H^{(2013)}}}$\,$\simeq$\,$12.6\,{\rm{N_H^{(2010)}}}$), the
2013 absorbing cloud should be larger in size by a factor of $\sim$\,$100$ 
with respect to the 2010 absorber. Since the 2010 absorber has an
estimated diameter larger than $5\times 10^{15}$\,cm, the 2013 cloud
would have a diameter of at least $5\times 10^{17}$\,cm. Even assuming
the highest possible orbital velocity for the range of radii
associated with the torus ($\sim$\,$1100$\,\kms ), a cloud of that size
would cover a negligibly small \xr\ emitting region for about
140~years.

The \xr\ history of \eso\ excludes that this is the case, as
variability associated with column density of the order of that
observed during the 2013 observation do occur on time--scales as short
as one month, as demonstrated by \cite{miniutti2014} using {\it Swift}
data. In order to produce such short time--scale variability, any
absorber must be smaller and denser by orders of magnitude, which
places the 2013 absorber much closer in, in a region of the parametr
space that is roughly consistent with the dust--free \gls{blr} (note,
however, that the 2013 absorber is not necessarily associated with a
cloud that produces the optical broad lines, as its ionisation
parameter is likely too high to produce the correct line
ratios). Moreover, if the 2013 absorber were associated with the dusty 
torus, UV variability between 2006 and 2013 would have been expected,
due to the dramatic increase in column density. However, as reported
by \cite{miniutti2014}, no decrease in UV fluxes was detected in 2013
despite the increase in the \xr\ column density by about one order of
magnitude, which strongly suggests that the 2013 c--phase is
dust--free and thus part of the \gls{blr} rather than of the dusty
torus. We then conclude that while the 2006 cold absorber is most
likely part of the same clumpy absorbing structure as in 2010 (at
spatial scales consistent with the dusty, clumpy torus), the 2013
absorber is more consistent with being associated with a
denser and dust--free clump of the \gls{blr}, in line with the ones
detected, for example, in NGC\,1365 \citep{risaliti2009mnras},
\swi\ \citep{sanfrutos2013} or Mrk\,335 \citep{Longinotti13}.

On the other hand, the h--phase has similar ionisation and column
density in all observations, but its outflow velocity appears to be
higher in the 2006 and 2013 \xmm\ observations than in the 2010
\chandra\ one. This may indicate that this highly ionised phase is
also somewhat clumpy, or that it is more efficiently accelerated when
the \xr\ continuum is suppressed by c--phase absorption (the
\chandra\ observation being the less absorbed).

\section{Summary and conclusions}

We present results from six observations of the polar--scattered
Seyfert\,1.2 galaxy \eso.  Four observations were taken by the
\gls{hetgs} on board \chandra\ within ten days in April 2010.  Two
more observations were taken with \xmm\ and are considered here for
comparison. The first one, obtained in 2006, is slightly more absorbed
than the \chandra\ one.  The second (2013) is much more heavily
absorbed (see \eg\ Fig.\,\ref{fig:xmmchanxmm}).

The high--resolution \chandra\ data are characterised by a rich set of
absorption lines that can be associated with three outflowing
absorbing gas phases with different ionisations. A highly ionised
phase (h--phase) is responsible for the Fe\,\textsc{xxv} and
Fe\,\textsc{xxvi} absorption lines and it also contributes to the
observed Si\,\textsc{xiv} and S\,\textsc{xvi} features. An
intermediate--ionisation zone (l--phase) is mostly revealed by
Si\,\textsc{xiii-xiv} and S\,\textsc{xv-xvi}, and contributes as well
at low energies where absorption due to Ne\,\textsc{x} and
Mg\,\textsc{xi-xii} is seen. A third, low--ionisation phase (c--phase)
is also detected and accounts for the Si\,\textsc{viii-x} lines while
contributing, together with the l--phase, at Ne\,\textsc{x} and
Mg\,\textsc{xi-xii} as well. The latter phase replaces the strictly
neutral absorber that is ubiquitously observed in Compton--thin
\gls{agn} at \xr\ energies. Here we show that this absorber is in
fact both ionised and outflowing. The three phases are outflowing with
velocities of the order of $1000-2000$\,\kms, and there is evidence
for the l--phase to be slightly faster than the other two. The
c--phase ionisation responds to luminosity variation on time--scales
as short as a few days, demonstrating that the gas is dense enough to
be in photoionisation equilibrium with the continuum on short
time--scales. Its clumpiness is suggested by the variation of its
column density between the 2006 and the 2010 observations \citep[as well as
by the presence of a hard scattered component which contributes
significantly to the \xr\ spectrum in heavily absorbed data sets,
see][]{miniutti2014}. On the other hand the warm\,/\,hot phases are
consistent with having the same ionisation and column density on both
short and long time--scales, suggesting that they are distributed in a
more homogeneous way.

We show that the data are consistent with three co--spatial phases
with similar outflow velocities and confined between the inner ($\sim$\,$
0.1$\,pc) and outer ($\sim$\,$1.6$\,pc) edges of the so--called clumpy,
dusty torus. Moreover, under this assumption, the three phases share
the same pressure. This calls for a rather natural scenario in which
relatively cold, dense clouds are pressure--confined by the more
homogeneous warm\,/\,hot phases. Such torus--scale outflow may well
represent the outer part of an outflow launched further in, which may
give rise to the full system of \gls{blr} and obscuring torus in \gls{agn}
replacing, with a wind solution, the classical structure of standard
Unification schemes \citep{emmering92, elvis2000, Elitzur06}.

The 2013 \xmm\ observation is much more heavily obscured by a gas
phase with one order of magnitude higher column density and ionisation
than in 2010. We show that this absorber is unlikely to be co--spatial
with the 2006 and 2010 c--phase at torus--like spatial scales, and
that it must be associated with a smaller, denser structure. This
places the 2013 c--phase within the dust--free \gls{blr}, although the
ionisation is likely too high to give rise to the observed optical
broad lines.
 
In our analysis we detect both the \gls{blr} and the outflowing torus
components, which implies that all structures are within observational
reach in \eso , possibly thanks to a particularly favourable viewing
angle of $\sim$\,$45^\circ$, intermediate between classical Seyfert\,1 and
Seyfert\,2 galaxies.

\section*{Acknowledgements}

This work is based on data obtained from the Chandra Data Archive and
the Chandra Source Catalogue. We made use of software provided by the
Chandra \xr\ Center (CXC). We also used observations obtained with
\xmm, an ESA science mission with instruments and contributions
directly funded by ESA Member States and NASA. Financial support for
this work was provided by the European Union through the Seventh
Framework Programme (FP7\,/\,2007--2013) under grant n. 312789. MS thanks
CSIC for support through a JAE--Predoc grant, and rejects public
cutbacks harmful for the common good (such as those against science).
YK aknowledges support from grant DGAPA PAIIPIT IN104215. 



\bibliographystyle{mnras}
\bibliography{references}

\begin{thebibliography}{}
\makeatletter
\relax
\def\mn@urlcharsother{\let\do\@makeother \do\$\do\&\do\#\do\^\do\_\do\%\do\~}
\def\mn@doi{\begingroup\mn@urlcharsother \@ifnextchar [ {\mn@doi@}
  {\mn@doi@[]}}
\def\mn@doi@[#1]#2{\def\@tempa{#1}\ifx\@tempa\@empty \href
  {http://dx.doi.org/#2} {doi:#2}\else \href {http://dx.doi.org/#2} {#1}\fi
  \endgroup}
\def\mn@eprint#1#2{\mn@eprint@#1:#2::\@nil}
\def\mn@eprint@arXiv#1{\href {http://arxiv.org/abs/#1} {{\tt arXiv:#1}}}
\def\mn@eprint@dblp#1{\href {http://dblp.uni-trier.de/rec/bibtex/#1.xml}
  {dblp:#1}}
\def\mn@eprint@#1:#2:#3:#4\@nil{\def\@tempa {#1}\def\@tempb {#2}\def\@tempc
  {#3}\ifx \@tempc \@empty \let \@tempc \@tempb \let \@tempb \@tempa \fi \ifx
  \@tempb \@empty \def\@tempb {arXiv}\fi \@ifundefined
  {mn@eprint@\@tempb}{\@tempb:\@tempc}{\expandafter \expandafter \csname
  mn@eprint@\@tempb\endcsname \expandafter{\@tempc}}}

\bibitem[\protect\citeauthoryear{Ag{\'{i}}s-Gonz{\'{a}}lez
  et~al.,}{Ag{\'{i}}s-Gonz{\'{a}}lez et~al.}{2014}]{agis2014}
Ag{\'{i}}s-Gonz{\'{a}}lez B.,  et~al., 2014, \mn@doi [\mnras]
  {10.1093/mnras/stu1358}, \href
  {http://adsabs.harvard.edu/abs/2014arXiv1407.1238A} {443, 2862}

\bibitem[\protect\citeauthoryear{Antonucci}{Antonucci}{1993}]{antonucci1993}
Antonucci R.,  1993, \mn@doi [\araa] {10.1146/annurev.aa.31.090193.002353},
  \href {http://adsabs.harvard.edu/abs/1993ARA%26A..31..473A} {31, 473}

\bibitem[\protect\citeauthoryear{Arnaud}{Arnaud}{1996}]{arnaud_xspec1996}
Arnaud K.~A.,  1996, in Astronomical Data Analysis Software and Systems V.
  p.~17

\bibitem[\protect\citeauthoryear{Bianchi, Piconcelli, Chiaberge, Bail{\'{o}}n,
  Matt  \& Fiore}{Bianchi et~al.}{2009}]{bianchi2009}
Bianchi S.,  Piconcelli E.,  Chiaberge M.,  Bail{\'{o}}n E.~J.,  Matt G.,
  Fiore F.,  2009, \mn@doi [\apj] {10.1088/0004-637X/695/1/781}, \href
  {http://adsabs.harvard.edu/abs/2009ApJ...695..781B} {695, 781}

\bibitem[\protect\citeauthoryear{Blustin, Page, Fuerst, Branduardi-Raymont  \&
  Ashton}{Blustin et~al.}{2005}]{Blustin05}
Blustin A.~J.,  Page M.~J.,  Fuerst S.~V.,  Branduardi-Raymont G.,   Ashton
  C.~E.,  2005, \mn@doi [\aap] {10.1051/0004-6361:20041775}, \href
  {http://adsabs.harvard.edu/abs/2005A%26A...431..111B} {431, 111}

\bibitem[\protect\citeauthoryear{Cash}{Cash}{1979}]{cash1979}
Cash W.,  1979, \mn@doi [\apj] {10.1086/156922}, \href
  {http://adsabs.harvard.edu/abs/1979ApJ...228..939C} {228, 939}

\bibitem[\protect\citeauthoryear{Dickens, Currie  \& Lucey}{Dickens
  et~al.}{1986}]{dickens1986}
Dickens R.~J.,  Currie M.~J.,   Lucey J.~R.,  1986, \mnras, \href
  {http://adsabs.harvard.edu/abs/1986MNRAS.220..679D} {220, 679}

\bibitem[\protect\citeauthoryear{Elitzur \& Shlosman}{Elitzur \&
  Shlosman}{2006}]{Elitzur06}
Elitzur M.,  Shlosman I.,  2006, \mn@doi [\apjl] {10.1086/508158}, \href
  {http://adsabs.harvard.edu/abs/2006ApJ...648L.101E} {648, L101}

\bibitem[\protect\citeauthoryear{Elvis}{Elvis}{2000}]{elvis2000}
Elvis M.,  2000, \mn@doi [\apj] {10.1086/317778}, \href
  {http://adsabs.harvard.edu/abs/2000ApJ...545...63E} {545, 63}

\bibitem[\protect\citeauthoryear{Elvis, Risaliti, Nicastro, Miller, Fiore  \&
  Puccetti}{Elvis et~al.}{2004}]{elvis2004}
Elvis M.,  Risaliti G.,  Nicastro F.,  Miller J.~M.,  Fiore F.,   Puccetti S.,
  2004, \mn@doi [\apjl] {10.1086/424380}, \href
  {http://adsabs.harvard.edu/abs/2004ApJ...615L..25E} {615, L25}

\bibitem[\protect\citeauthoryear{Emmering, Blandford  \& Shlosman}{Emmering
  et~al.}{1992}]{emmering92}
Emmering R.~T.,  Blandford R.~D.,   Shlosman I.,  1992, \mn@doi [\apj]
  {10.1086/170955}, \href {http://adsabs.harvard.edu/abs/1992ApJ...385..460E}
  {385, 460}

\bibitem[\protect\citeauthoryear{Fairall}{Fairall}{1986}]{fairall1986}
Fairall A.~P.,  1986, \mnras, \href
  {http://adsabs.harvard.edu/abs/1986MNRAS.218..453F} {218, 453}

\bibitem[\protect\citeauthoryear{Ferland et~al.,}{Ferland
  et~al.}{2013}]{ferland13_cloudy}
Ferland G.~J.,  et~al., 2013, \rmxaa, \href
  {http://adsabs.harvard.edu/abs/2013RMxAA..49..137F} {49, 137}

\bibitem[\protect\citeauthoryear{Gupta, Mathur, Krongold  \& Nicastro}{Gupta
  et~al.}{2013}]{Gupta13}
Gupta A.,  Mathur S.,  Krongold Y.,   Nicastro F.,  2013, \mn@doi [\apj]
  {10.1088/0004-637X/768/2/141}, \href
  {http://adsabs.harvard.edu/abs/2013ApJ...768..141G} {768, 141}

\bibitem[\protect\citeauthoryear{Jim{\'{e}}nez-Bail{\'{o}}n, Krongold, Bianchi,
  Matt, Santos-Lle{\'{o}}, Piconcelli  \& Schartel}{Jim{\'{e}}nez-Bail{\'{o}}n
  et~al.}{2008}]{bailon2008}
Jim{\'{e}}nez-Bail{\'{o}}n E.,  Krongold Y.,  Bianchi S.,  Matt G.,
  Santos-Lle{\'{o}} M.,  Piconcelli E.,   Schartel N.,  2008, \mn@doi [\mnras]
  {10.1111/j.1365-2966.2008.13976.x}, \href
  {http://adsabs.harvard.edu/abs/2008MNRAS.391.1359J} {391, 1359}

\bibitem[\protect\citeauthoryear{Kalberla, Burton, Hartmann, Arnal, Bajaja,
  Morras  \& P{\"{o}}ppel}{Kalberla et~al.}{2005}]{kalberla2005}
Kalberla P. M.~W.,  Burton W.~B.,  Hartmann D.,  Arnal E.~M.,  Bajaja E.,
  Morras R.,   P{\"{o}}ppel W. G.~L.,  2005, \mn@doi [\aap]
  {10.1051/0004-6361:20041864}, \href
  {http://adsabs.harvard.edu/abs/2005A%26A...440..775K} {440, 775}

\bibitem[\protect\citeauthoryear{Krolik \& Kriss}{Krolik \&
  Kriss}{2001}]{krolik2001}
Krolik J.~H.,  Kriss G.~A.,  2001, \mn@doi [\apj] {10.1086/323442}, \href
  {http://adsabs.harvard.edu/abs/2001ApJ...561..684K} {561, 684}

\bibitem[\protect\citeauthoryear{Krolik, McKee  \& Tarter}{Krolik
  et~al.}{1981}]{krolik81}
Krolik J.~H.,  McKee C.~F.,   Tarter C.~B.,  1981, \mn@doi [\apj]
  {10.1086/159303}, \href {http://adsabs.harvard.edu/abs/1981ApJ...249..422K}
  {249, 422}

\bibitem[\protect\citeauthoryear{Krongold, Nicastro, Brickhouse, Elvis, Liedahl
   \& Mathur}{Krongold et~al.}{2003}]{krongold_phase2003}
Krongold Y.,  Nicastro F.,  Brickhouse N.~S.,  Elvis M.,  Liedahl D.~A.,
  Mathur S.,  2003, \mn@doi [\apj] {10.1086/378639}, \href
  {http://adsabs.harvard.edu/abs/2003ApJ...597..832K} {597, 832}

\bibitem[\protect\citeauthoryear{Longinotti et~al.,}{Longinotti
  et~al.}{2013}]{Longinotti13}
Longinotti A.~L.,  et~al., 2013, \mn@doi [\apj] {10.1088/0004-637X/766/2/104},
  \href {http://adsabs.harvard.edu/abs/2013ApJ...766..104L} {766, 104}

\bibitem[\protect\citeauthoryear{Markowitz, Krumpe  \& Nikutta}{Markowitz
  et~al.}{2014}]{markowitz2014}
Markowitz A.~G.,  Krumpe M.,   Nikutta R.,  2014, \mn@doi [\mnras]
  {10.1093/mnras/stt2492}, \href
  {http://adsabs.harvard.edu/abs/2014MNRAS.439.1403M} {439, 1403}

\bibitem[\protect\citeauthoryear{Matt, Bianchi, Marinucci, Guainazzi, Iwasawa
  \& Jimenez~Bailon}{Matt et~al.}{2013}]{matt2013}
Matt G.,  Bianchi S.,  Marinucci A.,  Guainazzi M.,  Iwasawa K.,
  Jimenez~Bailon E.,  2013, \mn@doi [\aap] {10.1051/0004-6361/201321293}, \href
  {http://adsabs.harvard.edu/abs/2013A%26A...556A..91M} {556, A91}

\bibitem[\protect\citeauthoryear{Miniutti et~al.,}{Miniutti
  et~al.}{2014}]{miniutti2014}
Miniutti G.,  et~al., 2014, \mn@doi [\mnras] {10.1093/mnras/stt2005}, \href
  {http://adsabs.harvard.edu/abs/2014MNRAS.437.1776M} {437, 1776}

\bibitem[\protect\citeauthoryear{Nandra, O{'}Neill, George  \& Reeves}{Nandra
  et~al.}{2007}]{nandra_pexmon2007}
Nandra K.,  O{'}Neill P.~M.,  George I.~M.,   Reeves J.~N.,  2007, \mn@doi
  [\mnras] {10.1111/j.1365-2966.2007.12331.x}, \href
  {http://adsabs.harvard.edu/abs/2007MNRAS.382..194N} {382, 194}

\bibitem[\protect\citeauthoryear{Netzer}{Netzer}{2008}]{netzer2008}
Netzer H.,  2008, \mn@doi [New Astron. Rev.] {10.1016/j.newar.2008.06.009},
  \href {http://adsabs.harvard.edu/abs/2008NewAR..52..257N} {52, 257}

\bibitem[\protect\citeauthoryear{Nicastro, Fiore  \& Matt}{Nicastro
  et~al.}{1999}]{nicastro99}
Nicastro F.,  Fiore F.,   Matt G.,  1999, \mn@doi [\apj] {10.1086/307187},
  \href {http://adsabs.harvard.edu/abs/1999ApJ...517..108N} {517, 108}

\bibitem[\protect\citeauthoryear{Puccetti, Fiore, Risaliti, Capalbi, Elvis  \&
  Nicastro}{Puccetti et~al.}{2007}]{puccetti2007}
Puccetti S.,  Fiore F.,  Risaliti G.,  Capalbi M.,  Elvis M.,   Nicastro F.,
  2007, \mn@doi [\mnras] {10.1111/j.1365-2966.2007.11634.x}, \href
  {http://adsabs.harvard.edu/abs/2007MNRAS.377..607P} {377, 607}

\bibitem[\protect\citeauthoryear{Risaliti, Elvis  \& Nicastro}{Risaliti
  et~al.}{2002}]{risaliti_var2002}
Risaliti G.,  Elvis M.,   Nicastro F.,  2002, \mn@doi [\apj] {10.1086/324146},
  \href {http://adsabs.harvard.edu/abs/2002ApJ...571..234R} {571, 234}

\bibitem[\protect\citeauthoryear{Risaliti et~al.,}{Risaliti
  et~al.}{2009a}]{risaliti2009mnras}
Risaliti G.,  et~al., 2009a, \mn@doi [\mnras]
  {10.1111/j.1745-3933.2008.00580.x}, \href
  {http://adsabs.harvard.edu/abs/2009MNRAS.393L...1R} {393, L1}

\bibitem[\protect\citeauthoryear{Risaliti et~al.,}{Risaliti
  et~al.}{2009b}]{Risaliti09}
Risaliti G.,  et~al., 2009b, \mn@doi [\apj] {10.1088/0004-637X/696/1/160},
  \href {http://adsabs.harvard.edu/abs/2009ApJ...696..160R} {696, 160}

\bibitem[\protect\citeauthoryear{Sanfrutos, Miniutti,
  Ag{\'{i}}s-Gonz{\'{a}}lez, Fabian, Miller, Panessa  \& Zoghbi}{Sanfrutos
  et~al.}{2013}]{sanfrutos2013}
Sanfrutos M.,  Miniutti G.,  Ag{\'{i}}s-Gonz{\'{a}}lez B.,  Fabian A.~C.,
  Miller J.~M.,  Panessa F.,   Zoghbi A.,  2013, \mn@doi [\mnras]
  {10.1093/mnras/stt1675}, \href
  {http://adsabs.harvard.edu/abs/2013MNRAS.436.1588S} {436, 1588}

\bibitem[\protect\citeauthoryear{Schmid, Appenzeller  \& Burch}{Schmid
  et~al.}{2003}]{schmid2003}
Schmid H.~M.,  Appenzeller I.,   Burch U.,  2003, \mn@doi [\aap]
  {10.1051/0004-6361:20030558}, \href
  {http://adsabs.harvard.edu/abs/2003A%26A...404..505S} {404, 505}

\bibitem[\protect\citeauthoryear{Smith, Brickhouse, Liedahl  \& Raymond}{Smith
  et~al.}{2001}]{smith_apec2001}
Smith R.~K.,  Brickhouse N.~S.,  Liedahl D.~A.,   Raymond J.~C.,  2001, \mn@doi
  [\apj] {10.1086/322992}, \href
  {http://adsabs.harvard.edu/abs/2001ApJ...556L..91S} {556, L91}

\bibitem[\protect\citeauthoryear{Torresi, Grandi, Longinotti, Guainazzi,
  Palumbo, Tombesi  \& Nucita}{Torresi et~al.}{2010}]{Torresi10}
Torresi E.,  Grandi P.,  Longinotti A.~L.,  Guainazzi M.,  Palumbo G. G.~C.,
  Tombesi F.,   Nucita A.,  2010, \mn@doi [\mnras]
  {10.1111/j.1745-3933.2009.00773.x}, \href
  {http://adsabs.harvard.edu/abs/2010MNRAS.401L..10T} {401, L10}

\bibitem[\protect\citeauthoryear{V{\'{e}}ron-Cetty \&
  V{\'{e}}ron}{V{\'{e}}ron-Cetty \& V{\'{e}}ron}{2006}]{veron2006}
V{\'{e}}ron-Cetty M.~P.,  V{\'{e}}ron P.,  2006, \mn@doi [\aap]
  {10.1051/0004-6361:20065177}, \href
  {http://adsabs.harvard.edu/abs/2006A%26A...455..773V} {455, 773}

\bibitem[\protect\citeauthoryear{Wang \& Zhang}{Wang \& Zhang}{2007}]{wang07}
Wang J.~M.,  Zhang E.~P.,  2007, \mn@doi [\apj] {10.1086/513685}, \href
  {http://adsabs.harvard.edu/abs/2007ApJ...660.1072W} {660, 1072}

\bibitem[\protect\citeauthoryear{Zhang, Ji, Kallman, Yao, Froning, Gu  \&
  Kriss}{Zhang et~al.}{2015}]{zhang15}
Zhang S.~N.,  Ji L.,  Kallman T.~R.,  Yao Y.~S.,  Froning C.~S.,  Gu Q.~S.,
  Kriss G.~A.,  2015, \mn@doi [\mnras] {10.1093/mnras/stu2594}, \href
  {http://adsabs.harvard.edu/abs/2015MNRAS.447.2671Z} {447, 2671}

\makeatother
\end{thebibliography}




\bsp	
\label{lastpage}
\end{document}